\documentclass[journal]{IEEEtran}
\usepackage{graphicx}
\usepackage{epstopdf}
\usepackage{float}
\usepackage{algorithmic}
\usepackage{algorithm, algorithmic}
\usepackage{array}
\usepackage{amsmath}
\usepackage{amssymb}
\usepackage{amsthm}
\usepackage{eqparbox}
\usepackage{fixltx2e}
\usepackage{xcolor}
\usepackage{makecell}
\usepackage{mathrsfs}
\usepackage{amsmath}
\usepackage{multicol}
\usepackage{flushend}

\usepackage{url}
\usepackage{cite}
\usepackage[justification=centering]{caption}
\usepackage{bm}
\usepackage{threeparttable}
\usepackage{subeqnarray}
\usepackage{cases}
\ifCLASSOPTIONcompsoc
\usepackage[caption=false,font=normalsize,labelfont=sf,textfont=sf]{subfig}
\else
\usepackage[caption=false,font=footnotesize]{subfig}
\fi

\allowdisplaybreaks[4]

\newtheorem{mypro}{Proposition}

\newcommand{\diag}{\mathop{\mathrm{diag}}}



\begin{document}

\title{Joint Beamforming and Reflecting Design in Reconfigurable Intelligent Surface-Aided Multi-User Communication Systems }

\author{Xiaoyan Ma,~\IEEEmembership{Student Member, IEEE,}
Shuaishuai~Guo,~\IEEEmembership{Member, IEEE,}
        Haixia~Zhang,~\IEEEmembership{Senior Member, IEEE,}~ Yuguang Fang,~\IEEEmembership{Fellow, IEEE,}
and  Dongfeng~Yuan,~\IEEEmembership{Senior Member, IEEE}
\thanks{The work presented in this paper was supported in part by the
	Project of International Cooperation and Exchanges NSFC under Grant No. 61860206005 and in part by National Natural Science Foundation of China under Grant No. 61671278 and 61801266. (\emph{Corresponding author: Haixia Zhang}.)
}

\thanks{X. Ma  and D. Yuan are with Shandong Provincial Key Laboratory of Wireless Communication Technologies and School of Information Science and Engineering, Shandong University, Qingdao 266237, China (email: maxiaoyan06@mail.sdu.edu.cn; dfyuan@sdu.edu.cn).}
\thanks{S. Guo, and H. Zhang are with Shandong Provincial Key Laboratory of Wireless Communication Technologies and School of Control Science and Engineering, Shandong University, Jinan 250061, China (email: shuaishuai\textunderscore guo@sdu.edu.cn; haixia.zhang@sdu.edu.cn).}
\thanks{Y. Fang is  with the Computer, Electrical and
Mathematical Science and Engineering Division,  Department of Electrical and Computer Engineering
University of Florida, Gainesville, FL, USA (email: fang@ece.ufl.edu).}

}

\markboth{}%
{Shell \MakeLowercase{\textit{et al.}}: Bare Demo of IEEEtran.cls for IEEE Journals}
\maketitle
\IEEEpeerreviewmaketitle

\begin{abstract}
  Reconfigurable intelligent surface (RIS) provides a promising way to build the programmable wireless transmission environments in the future. Owing to the large number of reflecting elements used at the RIS, joint optimization for the active beamforming at the transmitter and the passive reflector at the RIS is usually complicated and time-consuming. To address this problem,  this paper proposes a low-complexity joint beamforming and reflecting algorithm based on fractional programing (FP). Specifically, we first consider a RIS-aided multi-user communication system with perfect channel state information (CSI) and formulate an optimization problem to maximize the sum rate of all users. Since the problem is nonconvex, we decompose the original problem into three disjoint subproblems. By introducing favorable auxiliary variables, we derive the closed-form expressions of the beamforming vectors and reflecting matrix in each subproblem, leading to a joint beamforming and reflecting algorithm with low complexity. We then extend our approach to handle the case when transmitter-RIS and RIS-receiver channels are not perfect and develop corresponding low-complexity joint beamforming and reflecting algorithm with practical channel estimation. Simulation results have verified the effectiveness of the proposed algorithms as compared to various benchmark schemes.
\end{abstract}	

	
\section{Introduction}

\IEEEPARstart{R}{econfigurable} intelligent surface (RIS) has emerged as a promising research topic in wireless communications \cite{Wu_Towards_smart,smart_DiRendo,Di_Renzo_Access}.
Specifically, RIS is a planar surface comprising a large number of low-cost passive reflecting elements, each of which can induce a phase shift to the incident signals \cite{124,125,126,127}. By smartly tuning the phase shifts of all reflecting elements according to the dynamic wireless channel, the reflected signals can be constructively added to those from other links to boost the desired signal power, thus significantly enhancing the communication performance without the need of deploying additional active base stations (BSs) or relays. Besides, from the implementation perspective,  RISs possess appealing features such as low profile and lightweight, thus can be easily mounted on walls or ceilings\cite{reviewer13}. Since RISs are complementary devices in wireless networks, deploying them in existing wireless systems does not require to change their standards and hardware, while only necessary modification of the communication protocols suffices. Furthermore, RISs are usually much cheaper than active BSs/relays, which can be easily deployed rapidly \cite{Beamforming_WQZR,compare_2}. Owing to these advantages, RISs have attracted intensive attention and many research works have been done, including high-layer protocol design \cite{Protocol1,Protocol2,Protocol3}, capacity analysis \cite{CapacityMax2,CapacityMax3,CapacityMax1}, energy/spectral efficiency maximization \cite{efficiency1,SYM2,reviewer11}, physical layer security design \cite{PHYsecurity1,PHYsecurity2,PHYsecurity3}, and modulation and coding \cite{Prof.Guo_Modulation,index-modulation2}.  Comprehensive surveys of these works on RISs can be found in \cite{DiRenzo_202004,reviewer22}.

New challenges also arise in the design and implementation of RIS-aided wireless
systems. How to jointly optimize the RIS reflection coefficients and the transmitter beamforming matrix to maximally reap the RIS performance gains is one of the most crucial problems\cite{SYM1,SZ_Chen}. Besides, as discussed in \cite{editor1}, the number of RIS reflecting elements can be large, which indicates that the joint optimization of the active beamforming at the transmitter and the passive reflecting at the RIS can be highly complicated and time-consuming. To address this problem, this paper targets at a low-complexity  joint beamforming and reflecting design for RIS-aided multi-user communication systems. Moreover, considering that RIS channel estimation (CE) are usually complicated, therefore, perfect channel state information (CSI) is difficult to obtain, low-complexity joint beamforming and reflecting design with practical channel information is further investigated.

\subsection{Related Works}
The joint beamforming and reflecting design in RIS-aided communication systems has aroused much research interest recently. Alternatively optimizing the beamforming and the reflecting matrices is a common method to solve this multivariate problem. To clearly review the state-of-the-art, we classify the related works into two categories according to the application scenarios as follows.

\emph{1) RIS-Aided Point-to-Point MIMO Systems.}
Point-to-point multiple-input multiple-output (MIMO) system is one of the classical models that have been well investigated. It is important to explore how much performance gains the communication systems can obtain from the deployment of RIS. Ye, Guo and Alouni attempted to answer this question from the system reliability perspective. It is shown in \cite{ye2019joint} that RIS can greatly reduce the symbol error rate (SER) by carefully designing the beamforming and reflecting matrices. In \cite{ye2019joint}, the joint design of the beamforming and reflecting matrices is optimized by a gradient descent algorithm with much higher computational complexity. Different from \cite{ye2019joint}, Zhang and Zhang attempted to explore the performance improvement brought by RIS from the spectral efficiency perspective. In \cite{zhangshuowen2019capacity}, they alternatively optimized the beamforming and reflecting matrices for spectral efficiency maximization. The closed-form beamforming and reflecting matrices in each optimization step are derived, which can greatly reduce the computational complexity. Results on RIS-aided point-to-point communication systems can well demonstrate the impacts of RIS on the communication links, but cannot be extended to the situations with the coexistence of other users, i.e., the interference links. As a result, the joint design of the beamforming and reflecting cannot be directly used in RIS-aided multi-user communication systems.

\emph{2) RIS-Aided Multi-User Communication Systems.}
For RIS-aided multi-user systems, the joint optimization method based on semidefinite relaxation (SDR) is one of the most commonly used methods. In \cite{SDR1,13}, the transmit power of a BS is minimized by decomposing the joint optimization problem into two subproblems: one is the conventional power-minimization problem in MIMO systems, while the other is the RIS phase vector optimization problem. Then the phase optimization problem is solved by SDR with high computational complexity.
Chen {\sl et al.} maximized the minimum-secrecy-rate in a downlink multi-input single-output (MISO) broadcast system by jointly optimizing the beamformers at the transmitter and reflecting coefficients at the RIS, and proposed an algorithm based on the alternating optimization and the path-following algorithm to solve the problem \cite{security}.
In \cite{Admm}, Guo {\sl et al.} combined the fractional programming and alternating direction method of multipliers (ADMM) to maximize the weighted sum rate of all users at downlink transmissions in the RIS-aided multi-user communication systems.
Although the alternating optimization algorithms mentioned above achieve quite good performance, the main shortcoming lies in that the complexity of those algorithms is extremely high for large-sized RISs.
In \cite{14} and \cite{15}, the energy efficiency is maximized in RIS-aided communication systems by employing zero-forcing (ZF) precoding at the transmitter. Since ZF precoding completely cancels the inter-user interference, the power allocation at the BS and the phase optimization at the RIS can be well decoupled. However, the ZF precoding may  amplify the background noise as well, and the performance could be severely compromised when the channel is ill-conditioned. Moreover, the derivations are not applicable directly for other precoding schemes. Di {\sl et al.} in \cite{reviewer12} proposed a hybrid beamforming scheme by assuming that there is a limited number of discrete phase shifts at RIS. Huang {\sl et al.} proposed a novel method leveraging the deep reinforcement learning (DRL) to jointly design the beamforming and reflecting matrix simultaneously \cite{reviewer23}.

\subsection{Contributions}
In this paper, we propose a low-complexity joint beamforming and reflecting design method for RIS-aided multi-user communication systems. The main contributions are summarized as follows.
\begin{itemize}
\item  We first investigate a RIS-aided multi-user communication system with perfect separate CSIs and formulate an optimization problem to maximize the sum rate of all users, which is multivariate and non-convex. To solve this problem, we decompose the original problem into three disjoint subproblems by utilizing the novel FP transformations. By introducing favorable auxiliary variables, we derive the closed-form optimal expressions of the beamforming vectors and reflecting matrix. 

\item Based on the closed-form optimal expressions, we design a low-complexity joint beamforming and reflecting algorithm to maximize the sum rate of all users. Due to the closed-form optimal expressions used in each iteration, the proposed algorithm enjoys extreme low complexity compared with traditional optimization methods.
	
\item Considering perfect CSIs are not always available in practical systems, we futher modify the proposed algorithm to cope with estimated CSI, the effect of channel estimation error on the system performance is also investigated. Numerical results show that the proposed algorithm can be easily combined with existing CE methods and achieve better sum rate performance than other benchmark schemes.  
\end{itemize}

The remainder of this paper is organized as follows. The system model and problem formulation are introduced in Section II. The detailed derivations of the proposed low-complexity alternating algorithm are presented in Section III. In Section IV, the proposed low-complexity algorithm is further modified by taking the practical channel estimation into consideration. Numerical simulation results are presented in Section V to verify the effectiveness of the proposed algorithms. Finally, we conclude this paper in Section VI. 

\subsection{Notations}
The notations used in this paper are listed as follows. $\mathbb{E}\lbrace \cdot\rbrace$ denotes statistical expectation. $\mathcal{CN}(\mu,\sigma^{2})$ denotes the circularly symmetric complex Gaussian (CSCG) distribution with mean $\mu$ and variance $\sigma^{2}$.
$||\mathbf{w}||$ denotes the Euclidean norm. $||\mathbf{W}||_{F}$ denotes the Frobenius norm.
$\mathbf{I}_{M}$ represents the $M \times M$ identity matrix. $\mathbf{G}^{T}$ and $\mathbf{G}^{H}$ denote the transpose and conjugate transpose of the matrix $\mathbf{G}$, respectively.   $\mathbf{G}^{-1}$ represents the  inverse of $\mathbf{G}$ while $\mathbf{G}^{+}$ represents the pseudo-inverse of $\mathbf{G}$.
$\text{Tr} [\mathbf{G}]$ is the trace of the matrix.
$\diag(\bm{\phi})$ denotes the diagonal matrix of vector $\bm{\phi}$.
$u_{i,q}$ represents the element at row $i$ and column $q$ in matrix $\mathbf{U} $, $v_{n}$ represents the $n$-th elements of vector $\mathbf{V}$.
For any real number, $\sqrt{a}$ denotes the square root of $a$.
For any complex variable $x$, $|x|$ denotes the absolute value, $\angle x$ represents its phase angle, $x^{*}$ denotes its conjugate,  $\Re \lbrace x \rbrace$ and $\Im \lbrace x \rbrace$ represent its real part and imaginary part, respectively.

\section{System Model and Problem Formulation}
\subsection{System and Channel Model}
\begin{figure}[!tb]
	\centering
	\includegraphics[width=0.4\textwidth]{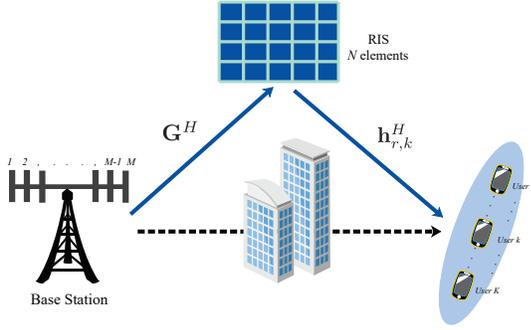}\\
	\caption{A RIS-aided multi-user communication system.}
	\label{fig:system_model}
\end{figure}

  This paper investigates a RIS-aided multi-user communication system as shown in Fig.  \ref{fig:system_model}, which consists of one BS equipped with $M$ antennas and $K$ signal antenna users. A RIS with $N$ reflecting elements is deployed between the BS and users to provide additional high-quality transmission links. In this paper, the direct links are weak due to the obstacles, thus BS chooses to use the RIS-aided links to improve the signal quality received at users.
 The equivalent channel from user $k$ to RIS and from RIS to BS are denoted by $ \mathbf{h}_{r,k} \in$$ \mathbb{C}^{N \times 1} $ and $ \mathbf{G} \in \mathbb{C}^{M \times N} $, respectively, where $ k=1,...,K $ represents the number of users.
  The RIS-aided link (i.e., user-RIS-BS link) is modeled as a concatenation of three components, i.e., the user-RIS link $\mathbf {h}_{r,k}$, RIS phase shift matrix (i.e., passive beamforming) $\mathbf{\Phi}\in \mathbb{C}^{N \times N}$, and RIS-BS link $\mathbf{G}$. $\mathbf{\Phi}=\diag (\bm{\phi}) $ is the diagonal matrix of the phase shift vector $\bm{\phi}$, where $\bm{\phi}=[\phi_{1},\phi_{2},...,\phi_{N}]^T \in \mathbb{C}^{N\times 1}$, $\phi_{n}=\beta_{n}e^{j\theta_n}$ with $n = 1,2,...,N$ denotes the phase and amplitude changes of each reflecting element introduced on the incident signals.
  $\theta_{n} \in [0,2\pi)$ and $\beta_{n} \in [0,1]$ denote the phase shift and the ON/OFF state of the $n$-th reflecting element at the RIS, respectively.
$\bm{\theta}=[\theta_{1}, \theta_{2},...,\theta_{N}]^{T}$ represents the phase angle vector of all reflecting elements.
In the subsequent development, we set $\beta_{n} =1$, which represents that all elements of the RIS are switched on to maximize the signal reflection.

It is assumed that the channels from BS to RIS and from RIS to users suffer from Rician fading. Similar  to \cite{channel_paras,channel_parameter}, we further assume that the antenna elements form a half-wavelength uniform linear array (ULA) configuration  at the AP and RIS, thus the channel $\mathbf{G}^H$ and $\mathbf{h}_{r,k}^H$ can be modeled by
 	\begin{small}
 	\begin{equation} \label{channel_model_G}
 		\mathbf{G}^H= d_{\mathbf{G}}\left(\sqrt{\frac{\mathcal{F}_{\mathbf{G}}}{\mathcal{F}_{\mathbf{G}}+1}}\bar{\mathbf{G}}^H  + \sqrt{\frac{1}{\mathcal{F}_{\mathbf{G}}+1}} \tilde{\mathbf{G}}^H\right),
 	\end{equation}
 \end{small}	
and
 	\begin{small}
 	\begin{equation} \label{channel_model_H}
 		\mathbf{h}_{r,k}^H= d_{r,k} \left(\sqrt{\frac{\mathcal{F}_{r,k}}{\mathcal{F}_{r,k}+1}}\bar{\mathbf{h}}_{r,k}^H + \sqrt{\frac{1}{\mathcal{F}_{r,k}+1}} \tilde{\mathbf{h}}_{r,k}^H\right) ,
 	\end{equation}
 \end{small}	
where $d_{\mathbf{G}}$ and $d_{r,k}$ denote the corresponding path loss at distance $d$; 
$\mathcal{F}_{\mathbf{G}}$ and $\mathcal{F}_{r,k}$ are the Rician factors of $\mathbf{G}^H$ and $\mathbf{h}_{r,k}^H$, respectively; $\bar{\mathbf{G}}^H$ and $\bar{\mathbf{h}}_{r,k}^H$ are the Line-of-Sight (LoS) components and can be expressed by the responses of the ULA. The array response of a $M$-element ULA is 
	\begin{equation}
		\begin{aligned} \label{ULA}
			\mathbf{a}_{M}(\varrho)=[1, e^{j2\pi\frac{d}{\mathcal{\lambda}}\sin \varrho},...,e^{j2\pi\frac{d}{\mathcal{\lambda}}(M-1)\sin \varrho} ],
		\end{aligned}
	\end{equation}
where $\varrho$ is the angular parameters. Under this condition, $\bar{\mathbf{G}}^H$ and $\bar{\mathbf{h}}_{r,k}^H$ are expressed as
\begin{equation}  
	\begin{aligned}\label{UL}
		\bar{\mathbf{G}}^H=\mathbf{a}_{N}^H(\eta)\mathbf{a}_{M}(\varrho),
	\end{aligned}
\end{equation}
and 
\begin{equation}
	\begin{aligned}
		\bar{\mathbf{h}}_{r,k}^H=\mathbf{a}_{N}(\iota_{k}),
	\end{aligned}
\end{equation}
where $\eta$, $\varrho$ and $\iota_k$ are the angular parameters,
$\tilde{\mathbf{G}}^H$ and $\tilde{\mathbf{h}}_{r,k}^H$ are the non-LoS (NLoS) components, whose elements are chosen from $ \mathcal{CN}(0,1) $. 

\subsection{Problem Formulation}
Let $x_{k}$ denote the transmitted signal from the BS to user $k$ with $\mathbb{E}\lbrace |x_k|^2  \rbrace=1, k=1,2,\cdots, K$.
Then, the transmitted signals for all users can be expressed as
\begin{eqnarray}
\begin{aligned}
\mathbf{x}=\sum_{k=1}^{K}\bm{ w}_{k}x_{k},
\end{aligned}
\end{eqnarray}
where $\bm{w}_{k} \in \mathbb{C}^{M \times 1}$ is the corresponding transmit beamforming vector for user $k$.
The signal received at user $k$ can be expressed as
\begin{eqnarray}
\begin{aligned}
y_{k}=\mathbf{h}_{r,k}^{H}\mathbf {\Phi} \mathbf {G}^{H} \mathbf{x} + n_{k},
\end{aligned}
\end{eqnarray}
where $n_{k} \sim \mathcal{CN}(0,\sigma_{0}^{2})$ denotes the complex additive white Gaussian noise (AWGN) at $k$-th user. The $k$-th user treats all the signals for other users (i.e., $x_{1},..., x_{k-1},x_{k+1},...,x_{K} $) as interference. Thus, the corresponding signal to interference plus noise ratio (SINR) at user $k$ can be expressed as
\begin{eqnarray}\label{Eq3}
\begin{aligned}\label{SINR_Ori}
\gamma_{k}=\frac{|\mathbf{h}_{r,k}^{H} \mathbf{\Phi} \mathbf{G}^{H} \bm{w}_{k}|^2}{\sum_{i=1,i\neq k}^{K}|\mathbf{h}_{r,k}^{H} \mathbf{\Phi} \mathbf{G}^{H} \bm{w}_{i}|^2+\sigma_{0}^{2}},
\end{aligned}
\end{eqnarray}
with
\begin{eqnarray}
\begin{aligned}
\sum_{k=1}^{K}||\bm{w}_{k}||^{2}\leq P_{T},
\end{aligned}
\end{eqnarray}
being the power constraint at BS, where $P_{T}$ represents the maximum transmit power. When BS uses all the power to transmit information for user $k$, i.e., $||\bm{w}_k||^2=P_T$, $\gamma_k$ can achieve the maximum value with $\gamma_k^{max}=\frac{||\mathbf{h}_{r,k}^{H} \mathbf{\Phi} \mathbf{G}^{H}||^2 P_T}{\sigma_0^2}$. Thus, the range of $\gamma_k$ is $0 \leq \gamma_k \leq \gamma_k^{max}$. \\

Let $\mathbf{W}=[\bm{w}_{1},\bm{w}_{2},...,\bm{w}_{K}]\in \mathbb{C}^{M \times K}$ denote the overall beamforming matrix at the BS. In this paper, we aim to maximize the sum rate of all users at the downlink transmission phase by jointly designing the active transmit beamforming matrix $\mathbf{W}$ at the BS and the passive reflecting matrix $\mathbf{\Phi}$ at the RIS, subject to the RIS phase shift constraint and the BS transmit power constraint.
Mathematically, the optimization problem is formulated as
\begin{align}
(P_{1}) \quad &\max \limits_{\mathbf{W},\mathbf{\Phi}} \quad f_{1}(\mathbf{W},\mathbf{\Phi})=\sum_{k=1}^{K} \log(1+\gamma_{k}),\\
&s.t.\quad \quad |\phi_{n}|=1, \quad n=1,2,...,N, \label{PSC}\\
& \quad\quad \quad \sum_{k=1}^{K}||\bm{w}_{k}||^{2}\leq P_{T}. \label{PC}
\end{align}
In ($P_{1}$), both the objective function $f_{1}(\mathbf{W},\mathbf{\Phi})$  and the phase shift constraints \eqref{PSC} are non-convex. Besides, the joint optimization of $\mathbf{W}$ and $\mathbf{\Phi}$ is deeply coupled. All these bring great difficulties to find the global optimal solution. In this paper, we attempt to find a low-complexity solution to ($P_{1}$) by applying the FP technique. Details are given in the next section.

\section{Joint Beamforming and Reflecting Design With Perfect CSI}

Problem ($P_{1}$) is to maximize a sum-of-log-of-ratio objective, i.e., $ \sum_{k=1}^{K} \log(1+\gamma_{k})$. It is a typical FP problem owing to the fractional expression of $\gamma_k$ in (\ref{SINR_Ori}).
In order to solve this non-convex multi-variable problem, we first introduce a series of auxiliary variables, and then transform ($P_1$) into a much low-complex problem by applying the FP techniques  \cite{method,method2}. It is written as
\begin{eqnarray} \label{ORP}
\begin{aligned}
(P_{1a}) \quad &\max \limits_{\bm{\alpha},\mathbf{W},\mathbf{\Phi}}\quad f_{1a}(\bm{\alpha},\mathbf{W},\mathbf{\Phi}),\\
& \quad s.t.\quad \quad |\phi_{n}|=1, \quad n=1,2,...,N,\\
& \quad \quad\quad \quad \sum_{k=1}^{K}||\bm{w}_{k}||^{2}\leq  P_{T},\\
&\quad \quad \quad \quad  \alpha_{k}\geq 0, \quad k=1,2,...,K.
\end{aligned}
\end{eqnarray}
where
\begin{eqnarray}
\begin{aligned}
f_{1a}(\bm{\alpha},\mathbf{W},\mathbf{\Phi})
=&\sum_{k=1}^{K} \log(1+\alpha_{k})-\sum_{k=1}^{K}\alpha_{k}\\
&+\sum_{k=1}^{K}\frac{(1+\alpha_{k}) \gamma_{k}}{1+\gamma_{k}},\\
\end{aligned}
\end{eqnarray}
and the parameters $\alpha_{1},\alpha_{2},...,\alpha_{K}$ are the introduced real auxiliary optimization variables for each user with $\bm{\alpha}=[\alpha_{1},\alpha_{2},...,\alpha_{K}]^{T}$ being the auxiliary variable vector. The original problem is transformed into an optimization problem with decision variables $\bm{\alpha},\mathbf{W}$, and $\mathbf{\Phi}$.  To efficiently solve the problem, we decouple $(P_{1a})$ into three disjoint subproblems and target to find the closed-form optimal solutions of each subproblem. Detailed information about how to alternatively optimize $\bm{\alpha}$, $\mathbf{W}$ and $\mathbf{\Phi}$ are given as follows.

\subsection{Closed-Form Alternating Optimization for Joint Beamforming and Reflecting Design}
 
\emph{1) Closed-Form Optimal Solution for $\bm{\alpha}$.}
First, we aim to find the optimal closed-form expression for $\bm{\alpha}$ with given $\mathbf{W}$ and $\mathbf{\Phi}$. Specifically, with fixed $\mathbf{W}$ and $\mathbf{\Phi}$, the objective function in $(P_{1a})$ can be seen as a non-constrained optimization problem only with respect to $\bm{\alpha}$.
\begin{mypro}
	 When $\mathbf{W}$ and $\mathbf{\Phi}$ are fixed, the optimal $\alpha_{k}$ can be expressed as
\begin{eqnarray}
\begin{aligned} \label{alpha}
\alpha_{k}^{op}=\gamma_{k}.
\end{aligned}
\end{eqnarray}
\end{mypro}
\begin{IEEEproof}
Taking the partial derivative of $ f_{1}(\bm{\alpha},\mathbf{W},\mathbf{\Phi})$ with respect to $\alpha_{k}$ yields
\begin{eqnarray}
\begin{aligned}\label{AP}
\frac{\partial f_{1}(\bm{\alpha},\mathbf{W},\mathbf{\Phi})}{\partial \alpha_{k}}
&=\frac{\gamma_{k}-\alpha_{k}}{(1+\gamma_{k})(1+\alpha_{k})}.
\end{aligned}
\end{eqnarray}
It is easy to verify that $\alpha_{k}^{op}=\gamma_{k}$ maximizes $f_{1}(\bm{\alpha},\mathbf{W},\mathbf{\Phi})$, since $\frac{\partial f_{1}(\bm{\alpha},\mathbf{W},\mathbf{\Phi})}{\partial \alpha_{k}}>0$ when $\alpha_k<\gamma_{k}$ and $\frac{\partial f_{1}(\bm{\alpha},\mathbf{W},\mathbf{\Phi})}{\partial \alpha_{k}}<0$ when $\alpha_k>\gamma_{k}$. Besides, substituting $\alpha_{k}^{op}=\gamma_{k}$ into $(P_{1a})$ yields
\begin{eqnarray} 
	\begin{aligned} \nonumber
	 f_{1a}(\bm{\alpha}^{op},\mathbf{W},\mathbf{\Phi})=\sum_{k=1}^{K} \log(1+\gamma_{k})=f_{1}(\mathbf{W},\mathbf{\Phi}),
	\end{aligned}
\end{eqnarray}
where $\bm{\alpha}^{op}=[\alpha_1^{op},\alpha_2^{op},\cdots,\alpha_K^{op}]^T$. This guarantees the equivalence between $(P_1)$ and $(P_{1a})$.
\end{IEEEproof}

\emph{2) Closed-Form Optimal Solution for Active Beamforming $\mathbf{W}$.}
When $\bm{\alpha}$ is fixed, the variable left for optimization in $(P_{1a})$ is $\gamma_{k}$ only. Thus the corresponding optimization problem can be rewritten as
\begin{eqnarray}\label{eq11}
\begin{aligned}
(P_{2}) \quad \max_{\mathbf{W},\mathbf{\Phi}}\quad &\sum_{k=1}^{K} \frac{\tilde{\alpha}_{k} \gamma_{k}}{1+\gamma_{k}},\\
 s.t. \quad &|\phi_{n}|=1, \quad n=1,2,...,N,\\
\quad &\sum_{k=1}^{K}||\bm{w}_{k}||^{2}\leq  P_{T},
\end{aligned}
\end{eqnarray}
where  $\tilde{\alpha}_{k}=1+\alpha_{k}$ for $k=1,2,...,K$.
For easy presentation, we adopt $\mathbf{h}_{k}^{H}=\mathbf{h}_{r,k}^{H} \mathbf{\Phi}  \mathbf{G}^{H}$ to present the equivalent channel. Thus, the SINR at user $k$  in (\ref{Eq3})  can be rewritten as
\begin{eqnarray}\label{eq12}
\begin{aligned}
\gamma_{k}=\frac{|\mathbf{h}_{k}^{H}\bm{ w}_{k}|^{2}}{\sum_{i=1,i\neq k}^{K}|\mathbf{h}_{k}^{H} \bm{w}_{i}|^{2}+\sigma_{0}^{2}}.
\end{aligned}
\end{eqnarray}
Substituting (\ref{eq12}) into (\ref{eq11}), we can reformulate the objective function in $(P_{2})$ to be

\begin{eqnarray}
\begin{aligned}
\sum_{k=1}^{K} \frac{\tilde{\alpha}_{k} \gamma_{k}}{1+\gamma_{k}}= \sum_{k=1}^{K} \frac{\tilde{\alpha}_{k}|\mathbf{h}_{k}^{H} \bm{w}_{k}|^{2}}{\sum_{i=1}^{K}|\mathbf{h}_{k}^{H}\bm{w}_{i}|^{2}+\sigma_{0}^{2}}.
\end{aligned}
\end{eqnarray}
Thus, given $\bm{\alpha}$ and $\mathbf{\Phi}$, optimizing $\mathbf{W}$ becomes

\begin{eqnarray}
\begin{aligned}
(P_{2a}) \quad &\max_{\mathbf{W}} \quad f_{2a}(\mathbf{W})\!=\!\sum_{k=1}^{K}\frac{\tilde{\alpha}_{k}|\mathbf{h}_{k}^{H} \bm{w}_{k}|^2}{\sum_{i=1}^{K}|\mathbf{h}_{k}^{H} \bm{w}_{i}|^{2}\!+\!\sigma_{0}^{2}},\\
&s.t. \quad \quad\sum_{k=1}^{K}||\bm{w}_{k}||^{2}\leq P_{T}.
\end{aligned}
\end{eqnarray}

It is obvious that $(P_{2a})$ is a multiple-ratio fractional programming problem. Using the quadratic transform proposed in \cite{method}, $(P_{2a})$ can be reformulated as a biconvex optimization problem with a new objective function as
\begin{eqnarray}
\begin{aligned}\label{15}
f_{2b}(\mathbf{W},\bm{\beta})=&\sum_{k=1}^{K}2\sqrt{\tilde{\alpha}_{k}} \Re\left\lbrace{\beta_{k}^{*}\mathbf{h}_{k}^{H}\bm{w}_{k}}\right\rbrace\\
&-\sum_{k=1}^{K}|\beta_{k}|^{2}\left(\sum_{i=1}^{K}|\mathbf{h}_{k}^{H}\bm{w}_{i}|^2+\sigma_{0}^{2}\right),
\end{aligned}
\end{eqnarray}
where $\beta_{1},\beta_{2},...,\beta_{K}$ are the newly introduced complex auxiliary variables and $\bm{\beta}=[\beta_{1},\beta_{2},...,\beta_{K}]^{T}$ is the auxiliary variable vector. Then, as proved in \cite{method} and \cite{method2}, solving $(P_{2a})$ with respect to $\mathbf{W}$ is equivalent to solving the following  biconvex problem with respect to $\mathbf{W}$ and $\bm{\beta}$,
\begin{eqnarray}
\begin{aligned}
(P_{2b})\quad &\max_{\mathbf{W},\bm{\beta}} \quad f_{2b}(\mathbf{W},\bm{\beta}),\\
& s.t.  \quad \sum_{k=1}^{K}||\bm{w}_{k}||^{2}\leq  P_{T}.
\end{aligned}
\end{eqnarray}
By analyzing $(P_{2b})$, we have the following proposition regarding to its solution.

\begin{mypro}
	The optimal beamforming vector $\bm{w}_{k}$ can be updated by
	\begin{eqnarray}\label{ww}
	\begin{aligned}
	\bm{w}_{k}^{op}=\sqrt{\tilde{\alpha}_{k}}\beta_{k}^{op}\left(\sum_{i=1}^{K}|\beta_{i}^{op}|^{2}\mathbf{h}_{i}\mathbf{h}_{i}^{H}+\lambda^{op}\mathbf{I}_{M}\right)^{-1}\mathbf{h}_{k},
	\end{aligned}
	\end{eqnarray}
	where
	\begin{eqnarray} \label{beta}
	\begin{aligned}
	\beta_{k}^{op}=\frac{\sqrt{\tilde{\alpha}_{k}}\mathbf{h}_{k}^{H}\bm{w}_{k}}{\sum_{i=1}^{K}|\mathbf{h}_{k}^{H}\bm{w}_{i}|^{2}+\sigma_{0}^{2}}.
	\end{aligned}
	\end{eqnarray}
In (\ref{ww}), $\lambda^{op}$ is the dual variable introduced for the power constraint $\sum_{k=1}^{K}||\bm{w}_{k}^{op}||^{2}=P_{T}$, which can be expressed as
	\begin{eqnarray}
	\begin{aligned}\label{lamm}
	\lambda^{op}=\min\left\lbrace \lambda^{op} \geq 0:\sum_{k=1}^{K}||\bm{w}_{k}^{op}||^{2}=P_{T} \right\rbrace.
	\end{aligned}
	\end{eqnarray}

\end{mypro}

\begin{IEEEproof}
First, when $\mathbf{W}$ is fixed, the partial gradient of $f_{2b}(\mathbf{W},\bm{\beta})$ with respect to $\beta_{k}$ can be expressed as
\begin{eqnarray}
\begin{aligned}
\frac{\partial f_{2b}(\mathbf{W},\bm{\beta})} {\partial \beta_{k}}\!=\!2\sqrt{\tilde{\alpha}_{k}}\mathbf{h}_{k}^{H}\mathbf{w}_{k}\!-\!2\beta_{k}\sum_{i=1}^{K}|\mathbf{h}_{k}^{H}\bm{w}_{i}|^{2}\!-\!2\beta_{k}\sigma_{0}^{2}.
\end{aligned}
\end{eqnarray}
By solving $\partial f_{2b}(\mathbf{W},\bm{\beta})/ \partial \beta_{k}=0$, the optimal FP auxiliary variable $\beta_{k}^{op}$ for fixed $\mathbf{W}$  can be derived as that in \eqref{beta}. Next,
with fixed $\beta_{k}^{op}$, $f_{2b}(\mathbf{W},\bm{\beta})$ can be rewritten as
	\begin{eqnarray}
	\begin{aligned}\label{20}
	f_{2b}(\mathbf{W},\bm{\beta}^{op})=&\sum_{k=1}^{K}\sqrt{\tilde{\alpha}_{k}}\left(\beta_{k}^{op^{*}}\mathbf{h}_{k}^{H}\bm{w}_{k}+\bm{w}_{k}^{H}\mathbf{h}_{k}\beta_{k}^{op}\right)\\
	&-\sum_{i=1}^{K}|\beta_{i}^{op}|^{2} \sum_{k=1}^{K}\bm{w}_{k}^{H}\mathbf{h}_{i}\mathbf{h}_{i}^{H}\bm{w}_{k}.
	\end{aligned}
	\end{eqnarray}
Considering the power constraint $\sum_{k=1}^{K}||\bm{w}_{k}||^{2}= P_{T}$, we write the Lagrangian function of $f_{2b}(\mathbf{W},\bm{\beta}^{op})$ as
	\begin{eqnarray}
	\begin{aligned}
	&f_{2b}^{(L)}(\lambda, \mathbf{W}, \bm{\beta}^{op})\!=\!\sum_{k=1}^{K}\sqrt{\tilde{\alpha}_{k}}\left(\beta_{k}^{op^{*}}\mathbf{h}_{k}^{H}\bm{w}_{k}\!+\!\bm{w}_{k}^{H}\mathbf{h}_{k}\beta_{k}^{op}\right)\\
	&\!-\!\sum_{i=1}^{K}|\beta_{i}^{op}|^{2} \sum_{k=1}^{K}\bm{w}_{k}^{H}\mathbf{h}_{i}\mathbf{h}_{i}^{H}\bm{w}_{k}
	\!-\!\lambda\left(\sum_{k=1}^{K}\bm{w}_{k}^{H}\bm{w}_{k}\!-\!P_{T}\right),
	\end{aligned}
	\end{eqnarray}
	where $\lambda$ is the dual variable introduced for the power constraint.
Taking the partial derivative of $f_{2b}^{(L)}(\lambda, \mathbf{W},\bm{\beta}^{op})$ with respect to $\bm{w}_{k}$, we have
	\begin{eqnarray}
	\begin{aligned} \label{Der}
	&\frac{\partial f_{2b}^{(L)}(\lambda, \mathbf{W},\bm{\beta}^{op})}{\partial \bm{w}_{k}}=\\
	&2\sqrt{\tilde{\alpha}_{k}}\mathbf{h}_{k}\beta_{k}^{op}\!-\!
	2\sum_{i=1}^{K}|\beta_{i}^{op}|^{2} \sum_{k=1}^{K}\mathbf{h}_{i}\mathbf{h}_{i}^{H}\bm{w}_{k}
	\!-\!2\lambda \mathbf{I}_{M} \bm{w}_{k}.	
	\end{aligned}
	\end{eqnarray}
Then, by solving $\partial f_{2b}^{(L)}(\lambda, \mathbf{W},\bm{\beta}^{op}) / \partial \bm{w}_{k}=0$, we can get the optimal expression of $\bm{w}_{k}^{op}$ in \eqref{ww}. For $\lambda^{op}$, taking the partial derivative of $f_{2b}^{(L)}(\lambda, \mathbf{W},\bm{\beta}^{op})$ with respect to $\lambda$, we have
\begin{eqnarray}
\begin{aligned}
&\frac{\partial f_{2b}^{(L)}(\lambda, \mathbf{W},\bm{\beta}^{op})}{\partial \lambda}=-\sum_{k=1}^{K}\bm{w}_{k}^{H}\bm{w}_{k}\!+\!P_{T}.
\end{aligned}
\end{eqnarray}
By solving the equation ${\partial f_{2b}^{(L)}(\lambda, \mathbf{W},\bm{\beta}^{op})}/{\partial \lambda}=0$, i.e., $\sum_{k=1}^{K}||\bm{w}_{k}^{op}||^{2}=P_{T}$, we can obtain all feasible solutions for $\lambda^{op}$. Among all solutions, the smallest number is chosen since $\lambda\geq 0$, $\sum_{k=1}^{K}\bm{w}_{k}^{H}\bm{w}_{k}-P_{T} \leq 0$,  and the dual function of $f_{2b}^{(L)}(\lambda,\mathbf{W},\bm{\beta}^{op})$, i.e., $\min \limits_{\lambda} \mathcal{G}(\lambda)	=\min \limits_{\lambda}  \max\limits_{\mathbf{W}, \bm{\beta}^{op}} f_{2b}^{(L)}(\lambda,\mathbf{W},\bm{\beta}^{op})$ is an increasing function with respect to $\lambda$, thus we need to select the smallest one, since minimizing $ \mathcal{G}(\lambda)$ is equals to maximizing $ f_{2b}^{(L)}(\lambda,\mathbf{W},\bm{\beta}^{op})$. Finally, we can get $\lambda^{op}$ in Eq. (\ref{lamm}).
\end{IEEEproof}

\emph{3) Closed-Form Optimal Solution for Passive Reflecting $\mathbf{\Phi}$.}
For fixed $\bm{\alpha}$, the objective function in $(P_{1a})$ is reduced to $(P_{2})$ in \eqref{eq11} as discussed above. At the reflecting optimization phase, we further rewrite the objective function of $(P_{2})$ as an equation of $\mathbf{\Phi}$, i.e.,
\begin{eqnarray}
\begin{aligned}
f_{3}(\mathbf{\Phi})&=\sum_{k=1}^{K} \frac{\tilde{\alpha}_{k} \gamma_{k}}{1+\gamma_{k}}\\
&=\sum_{k=1}^{K}\frac{\tilde{\alpha}_{k}|\mathbf{h}_{r,k}^{H}\mathbf{\Phi}\mathbf{G}^{H}\bm{w}_{k}|^{2}}{\sum_{i=1}^{K}|\mathbf{h}_{r,k}^{H}\mathbf{\Phi}\mathbf{G}^{H}\bm{w}_{i}+\sigma_{0}^{2}|}.
\end{aligned}
\end{eqnarray}
Recall that $\mathbf{\Phi}=\diag(\bm{\phi})$ with $\bm{\phi}=[\phi_{1},\phi_{2},...,\phi_{N}]^T$,  $f_{3}(\mathbf{\Phi})$ can be further expressed as a function of $\bm{\phi}$, i.e.,
\begin{eqnarray}
\begin{aligned}
f_{3}(\bm{\phi})
&=\sum_{k=1}^{K}\frac{\tilde{\alpha}_{k}|\bm{\phi}^{H}\diag (\mathbf{h}_{r,k}^{H})\mathbf{G}^{H}\bm{w}_{k}|^{2}}{\sum_{i=1}^{K}|\bm{\phi}^{H}\diag({\mathbf{h}_{r,k}^{H}})\mathbf{G}^{H}\bm{w}_{i}|^{2}+\sigma_{0}^{2}}\\
&=\sum_{k=1}^{K}\frac{\tilde{\alpha}_{k}|\bm{\phi}^{H} \bm{b}_{k,k}|^{2}}{\sum_{i=1}^{K}|\bm{\phi}^{H} \bm{b}_{i,k}|^{2}+\sigma_{0}^{2}},
\end{aligned}
\end{eqnarray}
where $\bm{b}_{i,k}=\diag(\mathbf{h}_{r,k}^{H})\mathbf{G}^{H}\bm{w}_{i}$ can be treated as the equivalent channel.
The corresponding optimization problem  can be transformed to be
\begin{eqnarray}\label{eqP3}
\begin{aligned}
(P_{3})\quad & \max_{\bm{\phi}} \quad  f_{3}(\bm{\phi})=\sum_{k=1}^{K}\frac{\tilde{\alpha}_{k}|\bm{\phi}^{H} \bm{b}_{k,k}|^{2}}{\sum_{i=1}^{K}|\bm{\phi}^{H} \bm{b}_{i,k}|^{2}+\sigma_{0}^{2}},\\
&  s.t. \quad \quad |\phi_{n}|=1, \quad n=1,2,...,N.
\end{aligned}
\end{eqnarray}
By solving ($P_{3}$), we obtain the following results.

\begin{mypro}
	The optimal $\phi_{n}, n=1,2,...,N$ in the reflecting vector $\bm{\phi}$ is updated by
	\begin{eqnarray}
	\begin{aligned}\label{phi}
	\phi_{n}^{op}=e^{j\theta_{n}^{op}},
	\end{aligned}
	\end{eqnarray}
with
\begin{eqnarray}\label{angle}
\begin{aligned}
\theta_{n}^{op}
&=\angle B_{2},\\
\end{aligned}
\end{eqnarray}
where  $B_{2}=v_{n}-\sum_{q=1,q\neq n}^{N} u_{n,q}\phi_{q}$,
$v_{n}$ is the $n$-th element in vector $\mathbf{V}=\sum_{k=1}^{K}\sqrt{\tilde{\alpha}_{k}}\varepsilon_{k}^{op{*}}\bm{b}_{k,k}$,
$u_{n,q}$ represents the element at row $i$ and column $q$ in matrix
$\mathbf{U}\!=\!\sum_{k=1}^{K}|\varepsilon_{k}^{op}|^{2}\sum_{i=1}^{K}\bm{b}_{i,k}\bm{b}_{i,k}^{H}$,
$\phi_{q}$ denotes the $q$-th element in $\bm{\phi}$ and  $\varepsilon_{k}^{op}$ is proved to be
\begin{eqnarray}
\begin{aligned} \label{varr}
\varepsilon_{k}^{op}=\frac{\sqrt{\tilde{\alpha}_{k}}\bm{\phi}^{H}\bm{b}_{i,k}}{\sum_{i=1}^{K}|\bm{\phi}^{H}\bm{b}_{i,k}|^{2}+\sigma_{0}^{2}}.
\end{aligned}
\end{eqnarray}
\end{mypro}

\begin{IEEEproof}
See Appendix A.
\end{IEEEproof}

\subsection{Low-complexity Alternating Algorithm for Joint Beamforming and Reflecting Design}
Based on the above analyses, we propose a low-complexity alternating algorithm to solve the joint beamforming and reflecting problem. 
The detailed steps of the proposed algorithm are listed in Algorithm 1. In Algorithm 1, $\bm{\alpha}$, $\mathbf{W}$ and $\mathbf{\Phi}$  are alternatively optimized with the closed-form expressions. Besides, we have the following conclusion.

\begin{mypro}
The proposed low-complexity alternating algorithm is guaranteed to converge.

\end{mypro}
\begin{IEEEproof}
See Appendix B.
\end{IEEEproof}

The computational complexity of Algorithm 1 mainly comes from the steps for the beamforming optimization and  the phase shift update. In the beamforming optimization step,  the complexity for calculating $\mathbf{W}$ is $\mathcal{O}(MK+K)$ and the complexity for updating $\bm{\alpha}$ is $\mathcal{O}(K)$. In the phase shift update step, for each user $k$, the optimal $\varepsilon_{k}^{op}$ is updated by \eqref{varr}, whose computational complexity can be estimated as $\mathcal{O}(K)$, the phase shift optimization of each reflecting element is designed according to \eqref{phi} at a computational complexity level of $\mathcal{O}(N)$ operations.
Summarizing all above, the aggregated complexity of Algorithm 1 can be expressed as $\mathcal{O}\left((N+MK+3K)I_{1}\right)$, with $I_1$ denoting the required number of outer iterations. From the analysis,  it is found that the computational complexity linearly increases with the number of reflecting elements, i.e., $N$. For more comparison, we list the complexity  and running time of different optimization algorithms for joint beamforming and reflecting design at Table \ref{table_complexity} in Section V. Observing the comparison, we find that the computational complexities of SDR based algorithm and ADMM based algorithm are much more sensitive to $N$ and not suitable for the applications with large $N$. We will give detailed performance comparisons about these algorithms in Section V.

\section{Proposed Joint Beamforming and Reflecting Design With Channel Estimation}
In this section, we will discuss the situation where perfect CSIs are not available, and we should seek the help from channel estimation (CE) to get CSI. We consider two types of CEs, separate CE and cascaded CE. For separate CE \cite{editor2,editor3,reviewer21},  the separate BS-RIS and RIS-user channels can be obtained. Under this circumstances, we can directly use the estimated BS-RIS and RIS-user channels to replace the perfect CSIs in the proposed algorithm. For cascaded CE, where only the cascaded BS-RIS-user channel can be obtained, we take the classical least-square (LS) CE for an example to show how the proposed algorithm works with the cascaded CSIs. We assume that the uplink and downlink channels have channel reciprocity. In order to acquire CSIs for the downlink transmission, uplink cascaded channel estimation is performed at the beginning of each time interval. However, since the RIS does not have any signal processing capabilities, only the cascaded user-RIS-BS channel can be estimated at the BS. During the CE phase, each user sends orthogonal pilot signals through the cascaded channel. Then, the estimated cascaded CSIs can be used for downlink beamforming and reflecting design. 
\begin{algorithm}[t]
	\caption{ Low-complexity optimization algorithm for joint beamforming and reflecting design \textbf{with perfect CSIs}}
	\label{alg:SA}
	\begin{algorithmic}[1]\label{alg:algorithm1}
		\STATE \textbf{Initialization:} Initialize $\mathbf{\Phi}^{0}$ and $\mathbf{W}^{0}$ that satisfy \eqref{PSC} and \eqref{PC}, respectively. Set the threshold value $\tau$ and the maximum number of iterations $I_m$. Iteration time $i=1$.\\
		\STATE \textbf{Step 1:}  Update $\bm{\alpha}^{i}$ according to (\ref{alpha}).
		\STATE \textbf{Step 2:}  Update $\bm{\beta}^{i}$ according to (\ref{beta}).
		\STATE \textbf{Step 3:}  Update $\mathbf{W}^{i}$ according to (\ref{ww}) and (\ref{lamm}).
	\STATE \textbf{Step 4:}  Update $\bm{\varepsilon}^{i}$ according to \eqref{varr}.
		\STATE \textbf{Step 5:}  Obtain $\bm{\phi}^{*}$ according to \eqref{phi} and \eqref{angle} .\\
		\IF      {$f_{3}(\bm{\phi}^{*})\ge f_{3}(\bm{\phi}^{i-1})$ }
		\STATE   $\bm{\phi}^{i}=\bm{\phi}^{*}$
		\ELSE
		\STATE   $\bm{\phi}^{i}=\bm{\phi}^{i-1}$
		\ENDIF\\
		\STATE Set $\mathbf{\Phi}^{i}=\diag (\bm{\phi}^{i})$.
		\STATE $i=i+1$.\\
		\textbf{UNTIL}\\ $f_{1a}(\bm{\alpha}^{i},\mathbf{W}^{i},\mathbf{\Phi}^{i})-f_{1a}(\bm{\alpha}^{i-1},\mathbf{W}^{i-1},\mathbf{\Phi}^{i-1})\leq \tau$ or the iteration time $i=I_m$.
	\end{algorithmic}
\end{algorithm}
\subsection{Cascaded Channel Estimation}
Taking user $k$ as an example and assuming that the transmit power at user $k$ is $P_{k}$, we can write the received pilot signals at the BS as
\begin{eqnarray}
\begin{aligned}
y_{k}&=P_{k}\mathbf{G}\mathbf{\Phi}_{l}\mathbf{h}_{r,k} s_{l}+n_{k}\\
&=P_{k}\mathbf{G} \diag(\mathbf{h}_{r,k})\bm{ \phi}_{l} s_{l}+n_{k}\\
&=P_{k}\mathbf{H}_{k} \bm{\phi}_{l} s_{l}+n_{k},
\end{aligned}
\end{eqnarray}
where $s_l$ represents the $l$-th pilot symbol satisfying $\mathbb{E}\lbrace |s_l|^2  \rbrace=1$, $\mathbf{\Phi}_{l}$ is the corresponding RIS phase shift matrix when transmitting the pilot signal $s_l$, $n_{k}$ denotes the AWGN noise in the uplink channel estimation phase with zero mean and $\sigma_k^2$ variance, and $\mathbf{H}_{k}=\mathbf{G} \diag(\mathbf{h}_{r,k})$ is the cascaded channel that needs to be estimated. Let $L$ denote the total pilot signal length used for channel estimation and  define $\mathbf{S}=P_{k}[s_{1},s_{2},...,s_{L}]^{T} \in \mathbb{C}^{L\times 1}$, we express the received signal during the channel estimation as
\begin{eqnarray}
\begin{aligned}
\mathbf{Y}_{k}&=\mathbf{H}_{k}[\bm{\phi}_{1},\bm{\phi}_{2},....,\bm{\phi}_{L}] \diag(\mathbf{S})+\mathbf{N}\\
&=\mathbf{H}_{k} \mathbf{X}+\mathbf{N},
\end{aligned}
\end{eqnarray}
where $\mathbf{\Psi}=[\bm{\phi}_{1},\bm{\phi}_{2},....,\bm{\phi}_{L}]\in \mathbb{C}^{N \times L}$ represents the equivalent RIS beamforming matrix during the channel estimation phase, $\mathbf{X}=\mathbf{ \Psi} \diag(\mathbf{S})$ can be seen as the equivalent training signals due to the fact that $\mathbf{ \Psi}$ and  $\diag(\mathbf{S})$ are known during the channel estimation, and $\mathbf{N}$ denotes the AWGN matrix. In this paper, we adopt a discrete Fourier transform (DFT) matrix as equivalent RIS beamforming matrix, i.e.,
$$
\mathbf{ \Psi}=\left[ \begin{matrix}
1        &  1         &  \cdots   &1\\
1        & \psi_{n}      &  \cdots   &\psi_{n}^{L-1}\\
\vdots   &\vdots      &  \ddots   &\vdots \\
1        &\psi_{n}^{L-1} &  \cdots   &\psi_{n}^{(N-1)(L-1)}
\end{matrix}  \right],
$$
with $\psi_{n}=e^{2\pi j/L}$. This is because a DFT matrix has the full rank and exhibits good performance for channel estimation \cite{DFT_M}.

Let $\mathbf{X}^{+}=\mathbf{X}^{H}(\mathbf{X}\mathbf{X}^{H})^{-1}$ represent the pseudo-inverse of $\mathbf{X}$ and a LS estimation for the cascaded user-RIS-BS channel can be expressed as $\hat{\mathbf{H}}_{k}=\mathbf{Y}_{k}\mathbf{X}^{+}$.
The corresponding channel estimation error can be calculated by
\begin{eqnarray}
\begin{aligned}
\mathbf {H}_{k}-\hat{\mathbf{H}}_{k}&=\mathbf {H}_{k}-\mathbf{Y}_{k}\mathbf{X}^{+}
=-\mathbf{N}\mathbf{X}^{+},
\end{aligned}
\end{eqnarray}
 the expectation of which can be computed as
\begin{eqnarray}
\begin{aligned}\label{m1}
\mathbb{E} \left\lbrace||\mathbf {H}_{k}-\hat{\mathbf{H}}_{k}||_{F}^{2}\right\rbrace&=\mathbb{E} \left \lbrace||\mathbf{N}\mathbf{X}^{+}||_F^2\right\rbrace\\
&=M \sigma_{n}^{2}\text{Tr}\left[(P_{k}\mathbf{\Psi}\mathbf{\Psi}^{H})^{-1}\right]\\
&=\frac{P_{k} M \sigma_{n}^{2} N}{L}.
\end{aligned}
\end{eqnarray}

\subsection{Proposed Low-complexity Alternating Optimization with Cascaded Channel Estimation}

According to the channel reciprocity, the cascaded BS-RIS-user channel estimation for downlink data transmission can be expressed by  $\hat{\mathbf{H}}_{k}^H$.
With  $\hat{\mathbf{H}}_{k}^{H}$, the joint beamforming and reflecting design based on perfect separate CSIs in Section III has to be revisited. Specifically, with the real cascaded downlink channel $\hat{\mathbf{H}}_{k}^{H}$ for each user $k$, the SINR in \eqref{SINR_Ori} can be modified to be
\begin{eqnarray}
\begin{aligned}
\gamma_{k}^{im}=\frac{|\bm{\phi}^{im^{H}}\hat{\mathbf{H}}_{k}^{H}\bm{w}_{k}^{im}|^2}
{\sum_{i=1,i\neq k}^{K}|\bm{\phi}^{im^{H}}\hat{\mathbf{H}}_{k}^{H} \bm{w}_{i}^{im}|^2+\sigma_{0}^{2}},
\end{aligned}
\end{eqnarray}
where $\hat{\mathbf{H}}_{k}^{H}=\diag(\mathbf{h}_{r,k}^{H}) \mathbf{G}^{H} $ represents the downlink cascaded channel for user $k$, $\bm{\phi}^{im}$ and $\bm{w}_{k}^{im}$ denote the RIS reflecting vector and BS beamforming vector designed based on the cascaded channel estimation $\hat{\mathbf{H}}_{k}^{H}$, respectively.

Then, the sum rate maximization problem with cascaded channel estimation can be reformulated as
\begin{align}
 (P_{1a}^{im}) \max \limits_{\bm{\alpha}^{im},\mathbf{W}^{im},\mathbf{\Phi}^{im}}
 \quad &f_{1}(\bm{\alpha}^{im}, \mathbf{W}^{im},\mathbf{\Phi}^{im}),\\
s.t. \quad\quad \quad &|\phi_{n}^{im}|=1, \quad n=1,2,...,N,\\
\quad \quad &\sum_{k=1}^{K}||\bm{w}_{k}^{im}||^{2}\leq P_{T},\\
\quad \quad &  0\leq \alpha_{k}^{im}, \quad k=1,2,...,K.
\end{align}
with
\begin{eqnarray}
\begin{aligned}
f_{1}(\bm{\alpha}^{im}, \mathbf{W}^{im},\mathbf{\Phi}^{im})
=&\sum_{k=1}^{K} \log(1+\alpha_{k}^{im})-\sum_{k=1}^{K}\alpha_{k}^{im}\\
&+\sum_{k=1}^{K}\frac{(1+\alpha_{k}^{im})\gamma_{k}^{im}}{1+\gamma_{k}^{im}},
\end{aligned}
\end{eqnarray}
and the parameters $\alpha_{1}^{im},\alpha_{2}^{im},...,\alpha_{k}^{im}$ are the introduced auxiliary optimization variables with $\bm{\alpha}^{im}=[\alpha_{1}^{im},\alpha_{2}^{im},...,\alpha_{K}^{im}]^{T}$ being the corresponding auxiliary variable vector,
$\mathbf{W}^{im}$ and $\mathbf{\Phi}^{im}$ represent the BS beamforming matrix and RIS reflecting matrix, respectively, $\phi_{n}^{im}$ is the phase shift of each reflecting element in $\bm{\phi}^{im}$ with $n=1,2,...,N$.

Different from the system design in Section III with perfect separate CSIs, the formulated three-variable joint beamforming and reflecting optimization problem in $(P_{1a}^{im})$ should be solved based on the estimated cascaded channel $\hat{\mathbf{H}}_{k}^{H}$. That is, the closed-form expressions of $\bm{\alpha}^{im}$, $\mathbf{W}^{im}$ and $\mathbf{\Phi}^{im}$ need to be found with respect to $\hat{\mathbf{H}}_{k}^{H}$. Through analyses, we have the following results, which are similar to Propositions 1-3.
\begin{mypro}
	The optimal expression of $\alpha_{k}^{im}$ with cascaded channel estimation can be updated by
	\begin{eqnarray}
	\begin{aligned} \label{alpha-im}
	\alpha_{k}^{im^{op}}=\gamma_{k}^{im}.
	\end{aligned}
	\end{eqnarray}
\end{mypro}
\begin{IEEEproof}
	$\alpha_{k}^{im^{op}}$ can be obtained by solving $\partial f_{1}(\bm{\alpha}^{im}, \mathbf{W}^{im},\mathbf{\Phi}^{im})/\partial \alpha_{k}^{im}=0$.

\end{IEEEproof}

Given $\bm{\alpha}^{im}$ and $\mathbf{\Phi}^{im}$, the optimization problem for finding the optimal $\mathbf{W}^{im}$ can be transformed by adopting the FP technique \cite{method} to be
\begin{eqnarray}
\begin{aligned}
(P_{2a}^{im})\quad \max_{\mathbf{W}^{im},\bm{\beta}^{im}} \quad &f_{2a}(\mathbf{W}^{im},\bm{\beta}^{im}),\\
s.t.  \quad\quad &\sum_{k=1}^{K}||\bm{w}_{k}^{im}||^{2}\leq  P_{T},
\end{aligned}
\end{eqnarray}
where $f_{2a}(\mathbf{W}^{im},\bm{\beta}^{im})$ can be expressed as
\begin{eqnarray}
\begin{aligned}
f_{2a}(\mathbf{W}^{im},\bm{\beta}^{im})\!=\!&\sum_{k=1}^{K}2\sqrt{\tilde{\alpha}_{k}^{im}} \Re\left\lbrace{\beta_{k}^{im^{*}}\mathbf{h}_{k}^{im^{H}}\bm{w}_{k}^{im}}\right\rbrace\\
&\!-\!\sum_{k=1}^{K}|\beta_{k}^{im}|^{2}\left(\sum_{i=1}^{K}|\mathbf{h}_{k}^{im^{H}}\bm{w}_{i}^{im}|^2\!+\!\sigma_{0}^{2}\right),
\end{aligned}
\end{eqnarray}
and parameters $\beta_{1}^{im},\beta_{2}^{im},...,\beta_{K}^{im}$ are the introduced auxiliary optimization variables with
$\bm{\beta}^{im}=[\beta_{1}^{im},\beta_{2}^{im},...,\beta_{K}^{im}]^{T}$ being the auxiliary variable vector, $\tilde{\alpha}_{k}^{im}=1+\alpha_{k}^{im}$, and $\mathbf{h}_{k}^{im}=\bm{\phi}^{im^{H}}  \hat{\mathbf{H}}_{k}^{H}$ is the equivalent channel represented by the cascaded channel estimation result $\hat{\mathbf{H}}_{k}^{H}$.

\begin{mypro}
	The optimal beamforming vector $\bm{w}_{k}^{im^{op}}$ is updated by
\begin{eqnarray}\label{ww-im}
\begin{aligned}
&\bm{w}_{k}^{im^{op}}\!=\! \\
&\sqrt{\tilde{\alpha}_{k}^{im}}\beta_{k}^{im^{op}}\left(\sum_{i=1}^{K}|\beta_{i}^{im^{op}}|^{2}\mathbf{h}_{i}^{im}\mathbf{h}_{i}^{im^{H}}\!+\!\lambda_{im}^{op}\mathbf{I}_{M}\right)^{-1}\mathbf{h}_{k}^{im},
\end{aligned}
\end{eqnarray}
where
\begin{eqnarray} \label{beta-im}
\begin{aligned}
\beta_{k}^{im^{op}}=\frac{\sqrt{\tilde{\alpha}_{k}^{im}}\mathbf{h}_{k}^{im^{H}}\bm{w}_{k}^{im}}{\sum_{i=1}^{K}|\mathbf{h}_{k}^{im^{H}}\bm{w}_{i}^{im}|^{2}+\sigma_{0}^{2}},
\end{aligned}
\end{eqnarray}
$\lambda_{im}^{op}$ is the dual variable introduced for the power
constraint, which satisfies $\sum_{k=1}^{K}||\bm{w}_{k}^{im^{op}}||^{2}=P_{T}$. That is, $\lambda_{im}^{op}$ can be expressed as
\begin{eqnarray}
\begin{aligned} \label{lamm-im}
\lambda_{im}^{op}=\min\left\lbrace \lambda_{im}^{op}\geq 0:\sum_{k=1}^{K}||\bm{w}_{k}^{im^{op}}||^{2}=P_{T} \right\rbrace.
\end{aligned}
\end{eqnarray}
\end{mypro}
\begin{IEEEproof}
Similar to the proof of Proposition 2, the optimal expression of $\beta_{k}^{im}$, i.e., $\beta_{k}^{im^{op}}$, can be obtained by setting $\partial f_{2a}(\mathbf{W}^{im},\bm{\beta}^{im}) / \partial \beta_{k}^{im}=0$. After $\beta_{k}^{im^{op}}$ is fixed, the optimal expression of $\bm{w}_{k}^{im^{op}}$ can also be obtained by setting $\partial f_{2a}(\mathbf{W}^{im},\bm{\beta}^{im}) / \partial \bm{w}_{k}^{im}=0$.
\end{IEEEproof}

At the RIS reflecting matrix optimization phase, since $\mathbf{\Phi}^{im}=\diag (\bm{\phi}^{im}) $ with $\bm{\phi}^{im}=[\phi_{1}^{im}, \phi_{2}^{im},...,\phi_{N}^{im}]^{T}$, the optimization problem can be rewritten as
\begin{eqnarray}
\begin{aligned}\label{50}
(P_{3a}^{im})\quad \quad \max_{\bm{\phi}^{im},\bm{\varepsilon}^{im}} \quad \quad &f_{3a}(\bm{\phi}^{im},\bm{\varepsilon}^{im}),\\
 s.t. \quad \quad \quad &|\phi_{n}^{im}|=1, \quad n=1,2,...,N, \quad
\end{aligned}
\end{eqnarray}
where
\begin{eqnarray}
\begin{aligned} \label{49}
f_{3a}(\bm{\phi}^{im},\bm{\varepsilon}^{im})
&\sum_{k=1}^{K}2\sqrt{\tilde{\alpha}_{k}^{im}} \Re\left\lbrace \varepsilon_{k}^{im^{*}}\bm{\phi}^{im^{H}}\bm{b}_{k,k}^{im}\right\rbrace\\
&\!-\!\sum_{k=1}^{K}|\varepsilon_{k}^{im}|^{2}\left(\sum_{i=1}^{K}|\bm{\phi}^{im^{H}}\bm{b}_{i,k}^{im}|\!+\!\sigma_{0}^{2}\right),\\
\end{aligned}
\end{eqnarray}
$\varepsilon_{1}^{im},\varepsilon_{2}^{im},...,\varepsilon_{N}^{im}$ are the introduced auxiliary optimization variables with
$\bm{\varepsilon}^{im}=[\varepsilon_{1}^{im},\varepsilon_{2}^{im},...,\varepsilon_{N}^{im}]^{T}$ being the auxiliary variable vector, and $\bm{b}_{i,k}^{im}=\hat{\mathbf{H}}_{k}^{H}\bm{w}_{i}^{im}$ denotes the equivalent channel represented by the cascaded channel estimation result $\hat{\mathbf{H}}_{k}^{H}$.
\begin{mypro}
The optimal  $\phi_{n}^{im}$ ($n=1,2,...,N$) in the reflecting vector $\bm{\phi}^{im}$ can be updated by
\begin{eqnarray}
\begin{aligned} \label{phi-im}
\phi_{n}^{im^{op}}=e^{j\theta_{n}^{im^{op}}},
\end{aligned}
\end{eqnarray}
with
\begin{eqnarray}\label{angle-im}
\begin{aligned}
\theta_{n}^{im^{op}}
&=\angle B_{2}^{im},
\end{aligned}
\end{eqnarray}	
where $B_{2}^{im}\!=\!v_{n}^{im}\!-\!\sum_{q\!=\!1,q\!\neq\! n}^{N} u_{n,q}^{im}\phi_{q}^{im}$,
$v_{n}^{im}$ is the $n$-th element in vector $\mathbf{V}^{im}\!=\!\sum_{k\!=\!1}^{K}\sqrt{\tilde{\alpha}_{k}^{im}}\varepsilon_{k}^{im^{op{*}}}\bm{b}_{k,k}^{im} $,
$u_{n,q}^{im}$ represents the element at row $i$ and column $q$ in matrix
$\mathbf{U}^{im}\!=\!\sum_{k=1}^{K}|\varepsilon_{k}^{im^{op}}|^{2}\sum_{i=1}^{K}\bm{b}_{i,k}^{im}\bm{b}_{i,k}^{im^{H}}$,
$\phi_{q}^{im}$ denotes the $q$-th element in $\bm{\phi}^{im}$, and
$\varepsilon_{k}^{im^{op}}$ is proved to be
\begin{eqnarray}
\begin{aligned}\label{var-im}
\varepsilon_{k}^{im^{op}}=\frac{\sqrt{\tilde{\alpha}_{k}^{im}}\bm{\phi}^{im^{H}}\bm{b}_{i,k}^{im}}{\sum_{i=1}^{K}|\bm{\phi}^{im^{H}}\bm{b}_{i,k}^{im}|^{2}+\sigma_{0}^{2}}.
\end{aligned}
\end{eqnarray}
\end{mypro}
\begin{IEEEproof}
See Appendix C.
\end{IEEEproof}

\subsection{Alternating Algorithm for Joint Beamforming and Reflecting Design With Cascaded CSIs}
Based on the above analysis, the proposed low-complexity joint beamforming and reflecting algorithm can be modified according to the cascaded CE results $\hat{\mathbf{H}}_{k}$.
 Similar to Algorithm 1, in Algorithm 2, three subproblems are alternatively updated with the closed-form optimal expressions of $\bm{\alpha}^{im^{op}}$, $\bm{w}_{k}^{im^{op}}$ and $\phi_{n}^{im^{op}}$. Specifically, $\bm{\alpha}^{im^{i}}$ is updated according to (\ref{alpha-im}), $\bm{\beta}^{im^{i}}$ is updated according to \eqref{beta-im}, $\mathbf{W}^{im^{i}}$ is updated according to (\ref{ww-im}) and (\ref{lamm-im}), $\bm{\phi}^{im^{*}}$ and $\bm{\varepsilon}^{im^{op}}$ are updated according to \eqref{phi-im}, (\ref{angle-im}) and \eqref{var-im}, respectively. Besides, we have the following conclusion.
\begin{mypro}
Algorithm 2 is guaranteed to converge.
\end{mypro}
\begin{IEEEproof}
Proposition 8 can be proved similarly to proposition 4, since the main difference between Algorithm 1 and Algorithm 2 is that the optimal  $\bm{\alpha}^{im^{op}}$, $\mathbf{w}_{k}^{im^{op}}$  and $\phi_{n}^{im^{op}}$ in Algorithm 2 are expressed by the cascaded channel estimation result $\hat{\mathbf{H}}_{k}^{H}$ instead of the perfect separate CSIs $\mathbf{G}^H$ and $\mathbf{h}_{r,k}^H$. 
\end{IEEEproof}
\begin{figure}[t]
	\vspace{-0.5cm}
	\centering
	\includegraphics[width=0.4\textwidth]{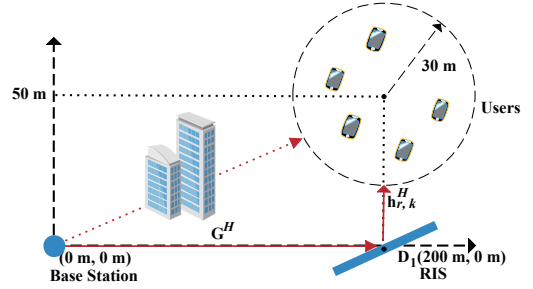}\\
	\caption{The simulated RIS-aided $K$-user communication system comprising of one $M$-antenna BS and one $N$-element RIS.}
	\label{fig:simulation_paras}
\end{figure}
\renewcommand\arraystretch{1.6}
\begin{table}[tb]
	\centering
	\caption{Simulation setups.}
	\label{table_par}
	\begin{threeparttable}
		\begin{tabular}{p{4.3cm}<{\centering}|p{3cm}<{\centering}}
			\hline
				\textbf{Parameters} & \textbf{Values} \\
			\hline
				BS location  & (0 m, 0 m)\\
			\hline
				RIS central location $D_1$  & (200 m, 0 m) \\
			\hline	
				Transmit power $P_T$ at the BS  &10 dBW	\\
			\hline	
				Large scale fading for $\mathbf{G}^H$ and $\mathbf{h}_{r,k}^H$&  35.6 + 22.0  $\log_{10}d$ dB\\
			\hline
				Rician factor for $\mathbf{G}^H$ and $\mathbf{h}_{r,k}^H$ &  10 dB\\
			\hline
		\end{tabular}	
	\end{threeparttable}
\end{table}
\renewcommand\arraystretch{1.6}
\begin{table*}[t]
	\centering
	\caption{Complexity comparison of different optimization algorithms for joint beamforming and reflecting design.}
	\label{table_complexity}
	\begin{threeparttable}
		\begin{tabular}{p{4.5cm}<{\centering}|p{5cm}<{\centering}|p{6cm}<{\centering}}
		
			\hline
			\textbf{Algorithm} & \textbf{Complexity} & \textbf{Running time for each iteration  when $ M=3, N=6, K=2$ (Unit: second)} \\
			
			\hline
			Proposed algorithm 	&  $\mathcal{O}\left((N+MK+3K)I_1 \right)$&0.0695\\
			
			\hline
	         ADMM based algorithm  & $\mathcal{O}\left((N^3+MK+2K)I_2\right)$& 2.3618\\	
			\hline
			MMSE + SDR based algorithm   & $\mathcal{O}\left((N^6+MK)I_3\right)$ &438.216\\
			\hline
		ZF + SDR based algorithm& $ \mathcal{O}\left( (N^6+MK)I_4  \right)$ & 435.558 \\
			\hline
		\end{tabular}			
		\begin{tablenotes}
			\footnotesize
			\item[*]  $I_n$ with $n=1,2,3,4$ is the number of outer iterations for different algorithms.
		\end{tablenotes}
	\end{threeparttable}
	\end{table*}
\section{Performance Evaluation}
This section provides numerical results to show the effectiveness of the proposed low-complexity algorithm for joint beamforming and reflecting in $(M,N,K)$ RIS-aided multi-user communication systems. This section is further divided into two subsections. Subsection V-A investigates the performance of the proposed design with perfect separate CSIs and Subsection V-B studies the performance of the proposed design with practical channel estimation. As discussed above, we investigate the performance of the RIS-aided $K$-user communication system as illustrated in Fig. \ref{fig:simulation_paras}. The $K$ single-antenna users are assumed to be uniformly and randomly distributed within a circle centered at ($200$ m, $50$ m) with a radius $30$ m. For clarity,  the system parameters are summarized in Table \ref{table_par}. The pathloss is set according to the 3GPP propagation environment described in \cite{3GPP}.
For comparison, we also simulate the performance of the following schemes:
\begin{itemize}
	\item   ADMM based scheme \cite{ADMM_Ori}.  It obtains the precoding vectors and phase shift matrices by solving non-convex problems in the ADMM method.
	\item MMSE/ZF + SDR based scheme \cite{SDR1}. Classical MMSE or ZF precoding schemes are directly adopted at the BS side, and the RIS phase shift matrix is obtained through the SDR method.
\end{itemize}
\subsection{Performance of the Proposed Design With Perfect CSIs}

\begin{figure}[!tb]
	\vspace{-0.63cm}
	\centering
	\includegraphics[width=0.4\textwidth]{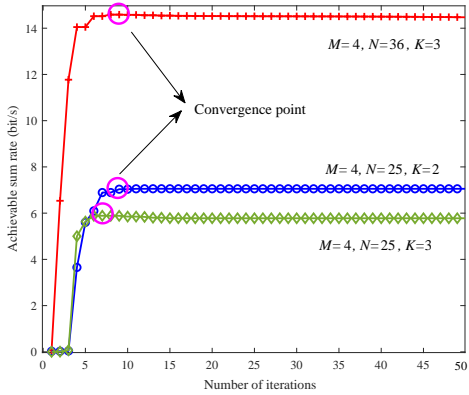}\\
	\caption{Convergence behavior of the proposed algorithm in $(4,25,3)$, $(4,25,2)$ and $(4,36,3)$ RIS-aided multi-user communication systems.}
	\label{fig:beamforming}
\end{figure}

First, we assume that BS knows all the CSIs of the involved channels, i.e., all separate CSIs, and investigate the convergence behavior and the performance in terms of sum rate of the proposed algorithm. Fig. \ref{fig:beamforming} shows the convergence behavior of the proposed low-complexity algorithm. It is observed that the proposed algorithm converges as stated in proposition 4 and converges fast in the depicted $(4,25,3)$, $(4,25,2)$ and $(4,36,3)$ RIS-aided multi-user communication systems. 

\begin{figure}[!tb]
	\centering
	\vspace{-0.15cm}
	\includegraphics[width=0.4\textwidth]{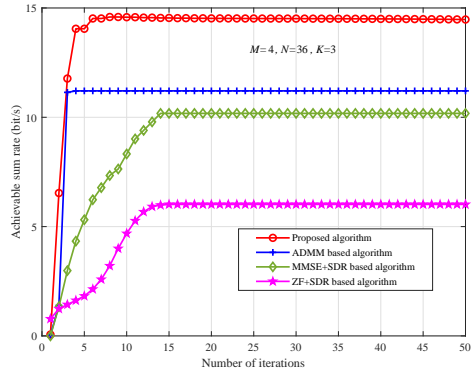}\\
	\caption{ Achievable sum rate of the proposed algorithm, ADMM based algorithm and MMSE/ZF+SDR algorithm in $(4,36,3)$ RIS-aided multi-user communication systems.}
	\label{fig:compare_one}
\end{figure}

To demonstrate the superiority of the proposed algorithm, we compare it with the ADMM based algorithm and traditional MMSE/ZF + SDR based algorithm in terms of the complexity. We list the complexity and running time of these algorithms in Table \ref{table_complexity}.  All simulations are performed on the same computer with a 3.2GHz Intel(R) Core(TM) i7-8700 CPU and 16GB RAM. In Fig. \ref{fig:compare_one}, we further show the convergence speed and the achievable sum rate of these algorithms. As it is shown, the proposed algorithm has lower complexity than the other algorithms in a $(4,36,3)$ RIS-aided multi-user system. Besides, the convergence speed of the proposed algorithm is much faster than those of the traditional MMSE/ZF + SDR based algorithm. The reason is that in each iteration, the algorithm is updated with the closed-form optimal solutions, which not only improves the sum rate performance of the algorithm, but also reduces the complexity and running time.

\begin{figure}[!tb]
	\centering
	\vspace{-0.6cm}
	\includegraphics[width=0.4\textwidth]{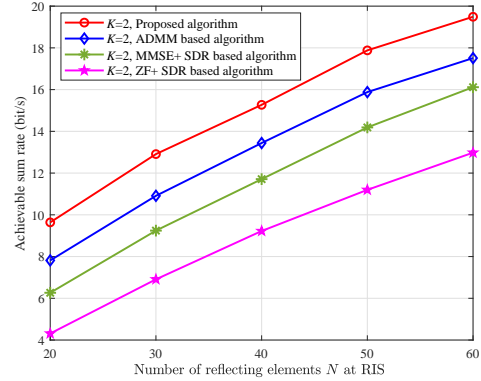}\\
	\caption{Achievable sum rate of the proposed algorithm, ADMM based algorithm and MMSE/ZF + SDR based algorithm when  the number of RIS reflecting elements increases, where $M=4$ and $K=2$ are fixed.}
	\label{fig:N20_60}
\end{figure}

For more comparison, we show the performance of the proposed algorithm, ADMM based algorithm and MMSE/ZF + SDR based algorithm in terms of achievable sum rate. As illustrated in Fig. \ref{fig:N20_60}, when $M=4$, $K=2$ are fixed, with the increase of RIS reflecting elements, the proposed algorithm always provides better system performance.

\begin{figure}[!tb]
	\centering
	\vspace{-0.54cm}
	\includegraphics[width=0.4\textwidth]{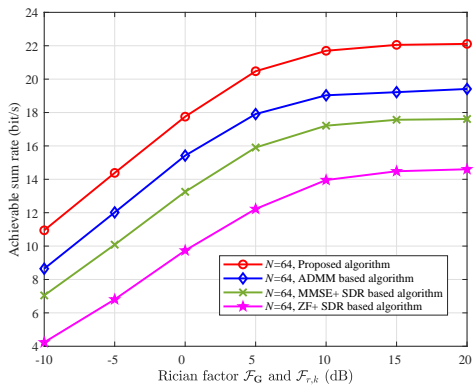}\\
	\caption{Achievable sum rate of the proposed algorithm, ADMM based algorithm, MMSE/ZF + SDR based algorithm when Rician factor changes ($\mathcal{F}_{\mathbf{G}}$ = $\mathcal{F}_{r,k}$), where $M=4$ and $K=2$.} 
	\label{fig:channel_Rician}
\end{figure}
\begin{figure}[!tb]
	\centering
	\vspace{-0.45cm}
	\includegraphics[width=0.4\textwidth]{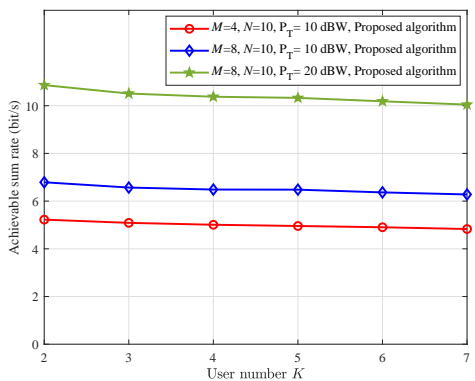}\\
	\caption{Achievable sum rate of the proposed algorithm when the number of users $K$ changes, $N=10$ is fixed.} 
	\label{fig:usernumber}
\end{figure}

Fig. \ref{fig:channel_Rician} shows how the sum rate is affected by the channel Rician factor. We assume that the channel from BS to RIS and from RIS to users follow Rician fading with Rician factors $\mathcal{F}_{\mathbf{G}}$ and $\mathcal{F}_{r,k}$. One can observe that with the increasing of Rician factors, the sum rate increases. This is because when the LoS component is getting stronger and stronger, the channel state is getting better, thus the sum rate increases accordingly. 

Fig. \ref{fig:usernumber} shows the relationship between the sum rate and the number of user $K$ for $N=10$. It is observed that the obtained sum rate decreases when $K$ increases. It implies that a larger number of users results in a lower average user rate due to the lower transmit power of each user and severer inter-user interference. Moreover, it is also found that a higher transmit power or larger number of BS antennas leads to a higher sum rate.

\subsection{Joint  Beamforming and Reflecting Design With Channel Estimation}

\begin{figure}[!tb]
	\centering
\vspace{-0.55cm}
	\includegraphics[width=0.4\textwidth]{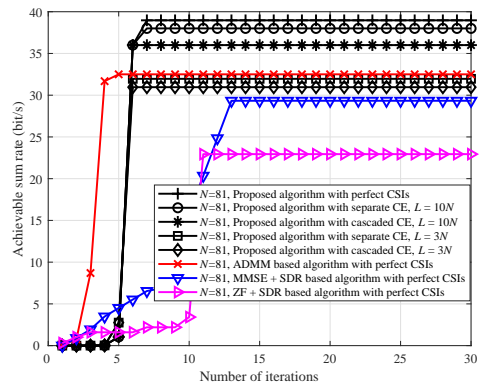}\\
	\caption{Achievable sum rate of the proposed algorithm, ADMM based algorithm, MMSE/ZF + SDR based algorithm under different CSIs, where $M=4$, $N=81$ and $K=2$.}
	\label{fig:COM_Pro_IM_AMDD_Per}
\end{figure}

We investigate the performance of the proposed algorithm with separate and cascaded CE. The separate CE method in \cite{editor3} and the cascaded LS CE method in Section IV-A are adopted. We assume that the channel coherent time is much larger than the pilot signal length. Simulation results are included in Fig. \ref{fig:COM_Pro_IM_AMDD_Per}, which show that the achievable sum rate increases as $L$ increases, because larger $L$ leads to accurate CSI. Also, the proposed algorithm combining with more advanced channel estimation methods, i.e., separate CE, achieves better performance. For comparison, we also include the performance of the ADMM based algorithm and MMSE/ZF + SDR based algorithm with perfect CSIs. It is shown that with sufficient long pilot signal length, the proposed algorithm with imperfect CSIs can outperform the ADMM based algorithm and MMSE/ZF + SDR based algorithm with perfect CSIs, which further verifies the effectiveness of the proposed algorithms. Besides, we can see that the convergence speed of the proposed algorithm for $N=81$ in Fig. \ref{fig:COM_Pro_IM_AMDD_Per} increases slightly compared with that for $N=36$ in Fig. \ref{fig:compare_one}, but they both can converge within around $10$ iterations. 

\begin{figure}[!tb]
	\centering
	\vspace{-0.07cm}
	\includegraphics[width=0.4\textwidth]{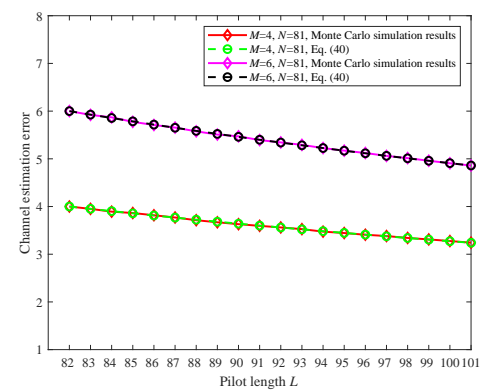}\\
	\caption{Channel estimation error versus pilot length $L$.} 
	\label{fig:channel_estimation_error}
\end{figure}
\begin{figure}[!tb]
	\centering
	\vspace{-0.4cm}
	\includegraphics[width=0.4\textwidth]{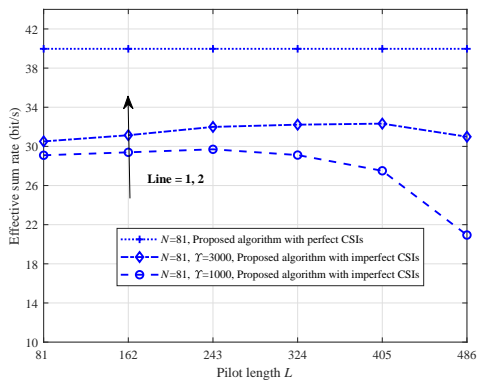}\\
	\caption{Effective sum rate of the proposed algorithm with different pilot signal length $L$ and time slot length $\Upsilon$, where $M=4$ and $K=2$.}
	\label{fig:Imperfect22}
\end{figure}

However, it is worth mentioned that the coherent time duration is limited in practical scenarios. Given a fixed coherent time, which is comparable to the pilot signal length, increasing the number of pilot signals no longer always increases the sum rate. This is because even though increasing the number of pilot signal length reduces the channel uncertainty, it also reduces the information transmission time. To demonstrate the impact of pilot signal length on the sum rate, we simulated the channel estimation accuracy and the results are shown in Fig. \ref{fig:channel_estimation_error}. It can be seen that the channel estimation error decreases when $L$ increases as analyzed in Eq. (\ref{m1}). We then simulate the performance of the proposed algorithm with different pilot signal length $L$ and time slot length $\Upsilon$ in Fig. \ref{fig:Imperfect22}. We define the effective sum rate $\tilde{R}=\frac{\Upsilon-L}{\Upsilon}\sum_{k=1}^{K} \log(1+\gamma_{k}^{im})$, where $L$ represents the time used for channel estimation in one time slot, and  $\Upsilon$ represents the total length of one time slot, thus $\Upsilon-L$ indicates the effective time for information transmission. Lines 1-2 first increases as $L$ increases since the channel estimation errors decreases as $L$ increases. Then, when $L$ keeps increasing, the effective sum rate will decrease since the time slots for information transmission $\Upsilon-L$ reduces and the gain brought by the improved channel accuracy cannot compensate for the loss caused by the reduction of $\Upsilon-L$.
\begin{figure}[!tb]
	\centering
	\vspace{-0.4cm}
	\includegraphics[width=0.4\textwidth]{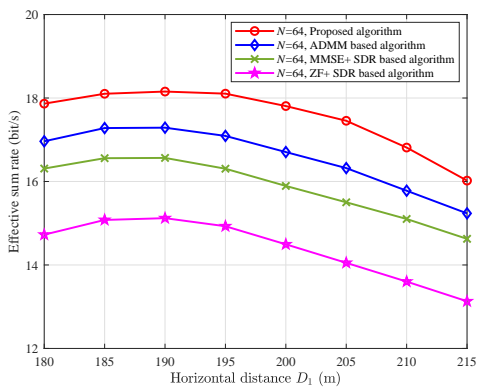}\\
	\caption{The sum rate of the proposed algorithm, ADMM based algorithm and MMSE/ZF + SDR based algorithm when the horizontal distance from BS to RIS changes, where $M=4$, $N=64$ and $K=3$.} 
	\label{fig:distance}
\end{figure}

Fig. \ref{fig:distance} illustrates the effective sum rate of all users when moving the RIS from (180 m, 0 m) to (215 m, 0 m). It can be seen that, when $D_1$ increases from 200 m to 215 m, the sum rate decreases. This is because both of the pass-loss of channel $\mathbf{G}^H$ and $\mathbf{h}_{r,k}^H$ increase. However, when decreasing $D_1$ from 200 m to 180 m, the sum rate first increases, and then decreases, and we get that the optimal location is $D_1$=190 m. This is because the path loss of the RIS-aided link is the product of the path losses of $\mathbf{G}^H$ and $\mathbf{h}_{r,k}^H$. Therefore, although the summation of the transmission distance decreases, the propagation condition might not become better.

\section{Conclusion}

In this paper, we have investigated a RIS-aided multi-user communication system where the direct links between the BS and the users are blocked and a RIS is deployed to provide additional high quality links to improve the system performance. The downlink sum rate optimization problem is formulated, which is a computationally intensive task due to the large number of RIS reflecting elements.
To address this non-convex multivariable problem, we have decoupled the original problem into three disjoint subproblems by utilizing fractional programming. By introducing auxiliary variables in each subproblem, we are able to obtain the closed-form expressions of the optimizing decision variables, and then, based on these, develop a low-complexity alternating optimization algorithm.  We have also shown that, the proposed algorithm enjoys extremely low complexity $\mathcal{O}\left((N+MK+3K)I_1 \right)$ with $N$ being the number of RIS reflecting elements.
Through extensive numerical analysis, our results have verified that the proposed algorithm  performs significantly better compared to various benchmark schemes.
\appendices
\section{Proof of Proposition 3}
\label{Appendix_A}
The optimization problem $(P_{3})$ is a multiple-ratio problem, which can be transformed into a biconvex optimization problem according to the quadratic transform proposed in \cite{method} and \cite{method2}. By introducing the new auxiliary variable  vector $\bm{\varepsilon}$, $f_{3}(\bm{\phi})$ can be rewritten as
\begin{eqnarray}
\begin{aligned}\label{f3a}
f_{3a}(\bm{\phi},\bm{\varepsilon})
=&\sum_{k=1}^{K}\sqrt{\tilde{\alpha}_{k}}\left(\varepsilon_{k}^{*}\bm{\phi}^{H}\bm{b}_{i,k}+\bm{b}_{i,k}^{H}\bm{\phi}\varepsilon_{k}\right)\\
&-\sum_{k=1}^{K}|\varepsilon_{k}|^{2}\left(\sum_{i=1}^{K}|\bm{\phi}^{H}\bm{b}_{i,k}|^{2}+\sigma_{0}^{2}\right),
\end{aligned}
\end{eqnarray}
where $\varepsilon_{1},\varepsilon_{2},...,\varepsilon_{N}$ are the introduced complex auxiliary variables and $\bm{\varepsilon}=[\varepsilon_{1},\varepsilon_{2},...,\varepsilon_{N}]^{T}$ is the auxiliary variable vector.
First, the optimal auxiliary variable $\varepsilon_{k}$, $k=1,2,...,N$ can be solved by setting $\partial f_{3a}(\bm{\phi},\bm{\varepsilon})/ \partial \varepsilon_{k}=0$. Taking partial derivative of $f_{3a}(\bm{\phi},\bm{\varepsilon})$ with respect to $\varepsilon_{k}$ yields
\begin{eqnarray}
	\begin{aligned}\nonumber
	\frac{\partial f_{3a}(\bm{\phi},\bm{\varepsilon})}{\partial \varepsilon_{k}}\!=\!
2\sqrt{\tilde{\alpha}_{k}}\bm{\phi}^{H}\bm{b}_{i,k}-2\varepsilon_{k}\sum_{i=1}^{K}|\bm{\phi}^{H}\bm{b}_{i,k}|^{2}-2\varepsilon_{k}\sigma_{0}^{2}.
	\end{aligned}
	\end{eqnarray}
	By setting $\partial f_{3a}(\bm{\phi},\bm{\varepsilon})/ \partial \varepsilon_{k}=0$, we can get the optimal $\varepsilon_{k}^{op}$ to be
\begin{eqnarray}\label{var}
\begin{aligned}
\varepsilon_{k}^{op}=\frac{\sqrt{\tilde{\alpha}_{k}}\bm{\phi}^{H}\bm{b}_{i,k}}{\sum_{i=1}^{K}|\bm{\phi}^{H}\bm{b}_{i,k}|^{2}+\sigma_{0}^{2}}.
\end{aligned}
\end{eqnarray}
 Substituting $\varepsilon_{k}^{op}$ into $f_{3a}(\bm{\phi},\bm{\varepsilon})$ yields
\begin{eqnarray}
\begin{aligned} \label{TR}
f_{3b}(\bm{\phi},\bm{\varepsilon}^{op})
=&\sum_{k=1}^{K}2\sqrt{\tilde{\alpha}_{k}}\Re\left\lbrace\varepsilon_{k}^{op^{*}}\bm{\phi}^{H}\bm{b}_{k,k}\right\rbrace\\ &\!-\!\sum_{k=1}^{K}|\varepsilon_{k}^{op}|^{2}\sum_{i=1}^{K}\bm{\phi}^{H}\bm{b}_{i,k}\bm{b}_{i,k}^{H}\bm{\phi}
\!-\!\sum_{k=1}^{K}|\varepsilon_{k}^{op}|^{2}\sigma_{0}^{2}\\
=&-\bm{\phi}^{H}\mathbf{U}\bm{\phi}+2\Re\left\lbrace \bm{\phi}^{H}\mathbf{V}\right\rbrace-\mathbf{C},
\end{aligned}
\end{eqnarray}
where
\begin{eqnarray}\label{27}
\begin{aligned}
&\mathbf{U}\!=\!\sum_{k=1}^{K}|\varepsilon_{k}^{op}|^{2}\sum_{i=1}^{K}\bm{b}_{i,k}\bm{b}_{i,k}^{H}, \quad \mathbf{V}\!=\!\sum_{k=1}^{K}\sqrt{\tilde{\alpha}_{k}}\varepsilon_{k}^{op{*}}\bm{b}_{k,k},\\
&\mathbf{C}\!=\!\sum_{k=1}^{K}|\varepsilon_{k}^{op}|^{2}\sigma_{0}^{2}.
\end{aligned}
\end{eqnarray}	

In order to obtain the closed-form optimal expression of $\bm{\phi}$, we define $u_{i,q}$ as the element at row $i$ and column $q$ in $\mathbf{U} \in \mathbb{C}^{N\times N}$, $v_{n}$ represents the $n$-th elements of vector $\mathbf{V} \in \mathbb{C}^{N \times 1}$. Recall that $\bm{\phi}=[\phi_{1},\phi_{2},...,\phi_{N}]^{T}$, in (\ref{TR}), the term $\bm{\phi}^{H}\mathbf{V}$ can be rewritten as
\begin{equation}
\begin{split}\label{TR1}
\bm{\phi}^{H}\mathbf{V}&=[\phi_{1}^{*},\phi_{2}^{*},...,\phi_{N}^{*}] \left[
\begin{matrix}
v_{1}\\
v_{2}\\
\vdots\\
v_{N}
\end{matrix}
\right]
=\sum_{i=1}^{N}\phi_{i}^{*}v_{i} \\
&=\phi_{n}^{*}v_{n}+\sum_{i=1,i\neq n}^{N}\phi_{i}^{*}v_{i}.
\end{split}
\end{equation}
The term $\bm{\phi}^{H}\mathbf{U}\bm{\phi}$ in (\ref{TR}) can be rewritten as
\begin{equation}
\begin{split} \label{TR2}
	\bm{\phi}^{H}\mathbf{U}\bm{\phi}\!=\!&[\phi_{1}^{*},\phi_{2}^{*},...,\phi_{N}^{*}]
	\left[
	\begin{matrix}
	u_{1,1}&u_{1,2}&\cdots&u_{1,N}\\
	u_{2,1}&u_{2,2}&\cdots&u_{2,N}\\
	\vdots&\vdots&\ddots&\vdots\\
	u_{N,1}&u_{N,2}&\cdots&u_{N,N}
	\end{matrix}
	\right]
	\left[
	\begin{matrix}
	\phi_{1}\\
	\phi_{2}\\
	\vdots\\
	\phi_{N}
	\end{matrix}
	\right] \\
	=&\sum_{i=1}^{n}\sum_{q=1}^{N}\phi_{i}^{*}u_{i,q}\phi_{j}\\
	=&\phi_{n}^{*}u_{n,n}\phi_{n}+2 \Re\left\lbrace\sum_{q=1,q\neq n}^{N}\phi_{n}^{*}u_{n,q}\phi_{q}\right\rbrace\\
	&+\sum_{i=1,i\neq n}^{n}\sum_{q=1,q\neq n}^{n} \phi_{i}^{*}u_{i,q}\phi_{q}.
\end{split}
\end{equation}

Substituting (\ref{TR1}) and (\ref{TR2}) into (\ref{TR}), $f_{3b}(\bm{\phi},\bm{\varepsilon}^{op})$ can be rewritten as
\begin{eqnarray}
\begin{aligned} \nonumber
f_{3c}(\phi_{n},\bm{\varepsilon}^{op})\!\!=\!&-\phi_{n}^{*}u_{n,n}\phi_{n}\!\!+\!\!2\Re\left\lbrace\phi_{n}^{*}v_{n}\!-\!\!\!\!\sum_{q=1,q\neq n}^{N}\phi_{n}^{*}u_{n,q}\phi_{q}\right\rbrace\\
&+2\Re\left\lbrace \sum_{i=1,i\neq n}^{N} \phi_{i}^{*}v_{i}\!-\!\!\!\!\!\!\sum_{i=1,i\neq n}^{N}\sum_{q=1,q\neq n}^{N}\phi_{i}^{*}u_{i,q}\phi_{q}\right\rbrace\\
=&-|\phi_{n}|^{2}B_{1}+2\Re\left\lbrace \phi_{n}^{*}B_{2}\right\rbrace+B_{3},
\end{aligned}
\end{eqnarray}
where
\begin{eqnarray}
\begin{aligned}
&B_{1}=u_{n,n}, \quad \quad B_{2}=v_{n}-\sum_{q=1,q\neq n}^{N} u_{n,q}\phi_{q},\\
&B_{3}=2\Re\left\lbrace \sum_{i=1,i\neq n}^{N} \phi_{i}^{*}v_{i}-\sum_{i=1,i\neq n}^{N}\sum_{q=1,q\neq n}^{N}\phi_{i}^{*}u_{i,q}\phi_{q}\right\rbrace.
\end{aligned}
\end{eqnarray}
Since $\phi_{n}=e^{j\theta_{n}}$, $|\phi_{n}|^{2}=1$. Finally, the optimization problem to find optimal $\mathbf{\Phi}$ given $\bm{\alpha}$ and $\mathbf{W}$ can be expressed as
\begin{eqnarray}
\begin{aligned} \label{SEE}
(P_{3c})\quad&\max_{\phi_{n}}\quad \quad f_{3c}(\phi_{n},\bm{\varepsilon}^{op})\!=\!-B_{1}\!+\!2\Re\left\lbrace \phi_{n}^{*} B_{2}\right\rbrace+B_{3},\\
& s.t. \quad \quad \quad |\phi_{n}|=1 \quad n=1,2,...,N.
\end{aligned}
\end{eqnarray}

From (\ref{SEE}), we can see that maximizing $f_{3c}(\phi_{n},\bm{\varepsilon}^{op})$ is equivalent to maximizing the term $2\Re\lbrace \phi_{n}^{*} B_{2}\rbrace$. Thus, the optimal solution of $\theta_{n}$ in \eqref{angle} can be obtained.\hfill $\blacksquare$

\section{Convergence analysis of Algorithm 1 }
\label{Appendix_B}
In Algorithm 1, Proposition 1 ensures that the update of $\bm{\alpha}$ always gives an improving value, that is, $f_{1a}(\bm{\alpha}^{i},\mathbf{W}^{i-1},\mathbf{\Phi}^{i-1})\ge f_{1a}(\bm{\alpha}^{i-1},\mathbf{W}^{i-1},\mathbf{\Phi}^{i-1})$ always satisfies in the $i$-th iteration.

For the BS beamforming optimization, when $\bm{\alpha}$ and $\mathbf{\Phi}$ are fixed, \eqref{15} is a biconvex problem related to $\bm{w}_{k}$ and $\beta_{k}$. For fixed $\mathbf{W}$, the optimal $\beta_{k}^{op}$ can be obtained by setting $\partial f_{2b}(\mathbf{W},\bm{\beta})/\partial \beta_{k}=0$.
For fixed $\beta_{k}$, the optimization problem to find $\mathbf{W}$ can be expressed as
\begin{eqnarray}
\begin{aligned}
\max_{\mathbf{W}} \quad f_{2b}(\mathbf{W})=&\sum_{k=1}^{K}\sqrt{\tilde{\alpha}_{k}}\left(\beta_{k}^{*}\mathbf{h}_{k}^{H}\bm{w}_{k}+\bm{w}_{k}^{H}\mathbf{h}_{k}\beta_{k}\right)\\
&-\sum_{i=1}^{K}|\beta_{i}|^{2} \sum_{k=1}^{K}\bm{w}_{k}^{H}\mathbf{h}_{i}\mathbf{h}_{i}^{H}\bm{w}_{k},\\
s.t.\quad\quad\quad\quad\quad\quad& \sum_{k=1}^{K}||\bm{w}_{k}||^{2}\leq  P_{T}.
\end{aligned}
\end{eqnarray}

Using the Lagrangian function, we can transform the above optimization problem into
\begin{eqnarray} \label{erci}
\begin{aligned}
f_{2b}^{(L)}(\mathbf{W})=&-\sum_{i=1}^{K}|\beta_{i}|^{2}\sum_{k=1}^{K}\bm{w}_{k}^{H}\mathbf{h}\mathbf{h}_{i}^{H}\bm{w}_{k}\\
&+\sum_{k=1}^{K}\sqrt{\tilde{\alpha}_{k}}\left(\beta_{k}^{*}\mathbf{h}_{k}^{H}\bm{w}_{k}+\bm{w}_{k}^{H}\mathbf{h}_{k}\beta_{k}\right)\\
&-\lambda \left(\sum_{k=1}^{K}\bm{w}_{k}^{H}\bm{w}_{k}-P_{T}\right).
\end{aligned}
\end{eqnarray}
By setting $\partial f_{2b}^{(L)}(\mathbf{W}) / \partial \bm{w}_{k}=0$, we can get the optimal expression of $\bm{w}_{k}$ in \eqref{ww}. By doing so, $f_{1a}(\bm{\alpha}^{i},\mathbf{W}^{i},\mathbf{\Phi}^{i-1})\ge f_{1a}(\bm{\alpha}^{i-1},\mathbf{W}^{i-1},\mathbf{\Phi}^{i-1})$ is guaranteed.

Finally, for fixed $\bm{\alpha}$ and $\mathbf{W}$, $f_{3a}(\bm{\phi},\bm{\varepsilon})$ in \eqref{f3a} is also a biconvex problem, we can get the optimal $\varepsilon_{k}^{op}$ by setting $\partial f_{3a}(\bm{\phi},\bm{\varepsilon})/ \partial \varepsilon_{k}=0$. Besides, we optimize the phase shift $\phi_{n}$ one by one according to \eqref{phi}. Based on all above analysis, we have $f_{3}(\bm{\phi}^{i})\ge f_{3}(\bm{\phi}^{i-1})$, which further guarantees that the objective function is non-decreasing. That is $f_{1a}(\bm{\alpha}^{i},\mathbf{W}^{i},\mathbf{\Phi}^{i})\ge f_{1a}(\bm{\alpha}^{i-1},\mathbf{W}^{i-1},\mathbf{\Phi}^{i-1})$ after the $i$-th iteration. Thus, the algorithm generates a non-decreasing sequence. 

To prove the convergence of Algorithm 1, we still need to show that the sequence generated from Algorithm 1 is upper bounded. As shown in \eqref{AP}, for $ 0 \leq \alpha_k \leq \gamma_k$ and $\bm{\alpha}=[\alpha_{1},\alpha_{2},...,\alpha_{K}]^{T}$, $f_{1a}(\bm{\alpha},\mathbf{W},\mathbf{\Phi})$ is an increasing function with respect to $\bm{\alpha}$. The maximum value of $f_{1a}(\bm{\alpha},\mathbf{W},\mathbf{\Phi})$ with respect to $\bm{\alpha}$ is achieved by setting $\alpha_k^{op} =\gamma_k$, leading to $f_{1a}(\bm{\alpha},\mathbf{W},\mathbf{\Phi})=f_{1}(\mathbf{W},\mathbf{\Phi})$. This shows that $f_{1a}(\bm{\alpha},\mathbf{W},\mathbf{\Phi})$ is upper-bounded by the original objective function $f_{1}(\mathbf{W},\mathbf{\Phi})$. Besides, since $\gamma_k$ ranges from $0$ to $\gamma_k^{max}$, $f_{1}(\mathbf{W},\mathbf{\Phi})$ must have a finite upper bound.  Thus, Algorithm 1 generates a bounded non-decreasing sequence and hence converges to at least a local optimum.

\hfill $\blacksquare$

\section{Proof of Proposition 7}

In \eqref{49}, the optimal $\varepsilon_{k}^{im}$, i.e., $\varepsilon_{k}^{im^{op}}$ can be obtained by setting $\partial f_{3a}(\bm{\phi}^{im},\bm{\varepsilon}^{im})/ \partial \varepsilon_{k}^{im}=0$ for $k=1,2,...,K$ as shown in \eqref{var-im}. After $\bm{\varepsilon}^{im^{op}}$ is obtained,
we further rewrite $f_{3a}(\bm{\phi}^{im},\bm{\varepsilon}^{im})$  as the following expression
\begin{eqnarray}
\begin{aligned} \label{65}
&f_{3b}(\bm{\phi}^{im},\bm{\varepsilon}^{im^{op}})\\
\!=\!&\sum_{k=1}^{K}2\sqrt{\tilde{\alpha}_{k}^{im}}\Re\left\lbrace\varepsilon_{k}^{im^{op^*}}\bm{\phi}^{im^{H}}\bm{b}_{k,k}^{im}\right\rbrace\\ &\!-\!\sum_{k=1}^{K}|\varepsilon_{k}^{im^{op}}|^{2}\sum_{i=1}^{K}\bm{\phi}^{im^{H}}\bm{b}_{i,k}^{im}\bm{b}_{i,k}^{im^{H}}\bm{\phi}^{im}
\!-\!\sum_{k=1}^{K}|\varepsilon_{k}^{im^{op}}|^{2}\sigma_{0}^{2}\\
&\!=\!-\bm{\phi}^{im^{H}}\mathbf{U}^{im}\bm{\phi}^{im}\!+\!2\Re\left\lbrace \bm{\phi}^{im^{H}}\mathbf{V}^{im}\right\rbrace-\mathbf{C}^{im},
\end{aligned}
\end{eqnarray}
where
\begin{eqnarray}
\begin{aligned} \nonumber &\mathbf{U}^{im}\!=\!\sum_{k=1}^{K}|\varepsilon_{k}^{im^{op}}|^{2}\sum_{i=1}^{K}\bm{b}_{i,k}^{im}\bm{b}_{i,k}^{im^{H}}, \mathbf{V}^{im}\!=\!\sum_{k=1}^{K}\sqrt{\tilde{\alpha}_{k}^{im}}\varepsilon_{k}^{im^{op^*}}\bm{b}_{k,k}^{im},\\
&\mathbf{C}^{im}=\sum_{k=1}^{K}|\varepsilon_{k}^{im^{op}}|^{2}\sigma_{0}^{2}.
\end{aligned}
\end{eqnarray}
The term $\bm{\phi}^{im^{H}}\mathbf{V}^{im}$ in \eqref{65} can be rewritten as
\begin{equation}
\begin{split} \label{62}
\bm{\phi}^{im^{H}}\mathbf{V}^{im}&
=\sum_{i=1}^{N}\phi_{i}^{im^{*}}v_{i}^{im} =\phi_{n}^{im^{*}}v_{n}^{im}+\sum_{i=1,i\neq n}^{N}\phi_{i}^{im^{*}}v_{i}^{im}.
\end{split}
\end{equation}
The term $\bm{\phi}^{im^{H}}\mathbf{U}^{im}\bm{\phi}^{im}$ in \eqref{65} can be rewritten as
\begin{equation}
\begin{split} \label{63}
\bm{\phi}^{im^{H}}\mathbf{U}^{im}\bm{\phi}^{im}
=&\sum_{i=1}^{n}\sum_{j=1}^{N}\phi_{i}^{im^{*}}u_{i,j}^{im}\phi_{j}^{im}\\
=&\phi_{n}^{im^{*}}u_{n,n}^{im}\phi_{n}^{im}\!+\!2 \Re\left\lbrace\sum_{j=1,j\neq n}^{N}\phi_{n}^{im^{*}}u_{n,j}^{im}\phi_{j}^{im}\right\rbrace\\
&+\sum_{i=1,i\neq n}^{n}\sum_{j=1,j\neq n}^{n} \phi_{i}^{im^{*}}u_{i,j}^{im}\phi_{j}^{im}.
\end{split}
\end{equation}
Using the simplified expressions in \eqref{62} and \eqref{63}, \eqref{65} can be rewritten as
\begin{eqnarray}
\begin{aligned} \nonumber
&f_{3c}(\phi_{n}^{im},\bm{\varepsilon}^{im^{op}})
\!=\!-|\phi_{n}^{im}|^{2}B_{1}^{im}\!+\!2\Re\left\lbrace \phi_{n}^{im^{*}}B_{2}^{im}\right\rbrace\!+\!B_{3}^{im},
\end{aligned}
\end{eqnarray}
where
\begin{eqnarray}
\begin{aligned}\nonumber
&B_{1}^{im}=u_{n,n}^{im}, \quad \quad B_{2}^{im}=v_{n}^{im}-\sum_{j=1,j\neq n}^{N} u_{n,j}^{im}\phi_{j}^{im},\\
&B_{3}^{im}\!=\!2\Re\left\lbrace \sum_{i=1,i\neq n}^{N} \phi_{i}^{im^{*}}v_{i}^{im}\!-\!\!\!\!\sum_{i=1,i\neq n}^{N}\sum_{j=1,j\neq n}^{N}\phi_{i}^{im^{*}}u_{i,j}^{im}\phi_{j}^{im}\right\rbrace.
\end{aligned}
\end{eqnarray}

Since $\phi_{n}^{im}=e^{j\theta_{n}^{im}}$, $|\phi_{n}^{im}|^{2}=1$. Finally, the optimization problem of $\mathbf{\Phi}^{im}$ when $\bm{\alpha}^{im}$ and $\mathbf{W}^{im}$ are fixed can be expressed as
\begin{eqnarray}
\begin{aligned}\label{FFin}
\max_{\bm{\phi}^{im}}\quad  &f_{3c}(\phi_{n}^{im},\bm{\varepsilon}^{im^{op}})\!=-B_{1}^{im}\!+\!2\Re\left\lbrace \phi_{n}^{im^{*}}B_{2}^{im}\right\rbrace\!+\!B_{3}^{im},\\
s.t. \quad  &\theta_{n}^{im} \in [0,2\pi) \quad n=1,2,...,N.
\end{aligned}
\end{eqnarray}

It can be seen from  (\ref{FFin}) that maximize $f_{3c}(\phi_{n}^{im},\bm{\varepsilon}^{im^{op}})$ is equivalent to maximize $2\Re\lbrace \phi_{n}^{im^{*}} B_{2}^{im}\rbrace$. Thus, the optimal solution of $\theta_{n}^{im}$ can be obtained by setting $\theta_{n}^{im^{op}}=\angle B_{2}^{im}$ as shown in \eqref{angle-im}.	
\hfill $\blacksquare$


\end{document}